\documentclass[aps,prb,twocolumn,superscriptaddress,showpacs,amsmath,amssymb,floatfix]{revtex4-1}
\usepackage{amsfonts}
\usepackage{graphicx}
\usepackage{dcolumn}
\usepackage{bm}
\usepackage[colorlinks,bookmarks=false,citecolor=blue,linkcolor=blue]{hyperref}

\newcommand{\ba}{\begin{array}}
\newcommand{\ea}{\end{array}}
\newcommand{\be}{\begin{equation}}
\newcommand{\ee}{\end{equation}}
\newcommand{\bea}{\begin{eqnarray}}
\newcommand{\eea}{\end{eqnarray}}
\newcommand{\bi}{\begin{itemize}}
\newcommand{\ei}{\end{itemize}}

\newcommand{\bmth}[1]{\mbox{\boldmath$#1$}}
\newcommand{\bmths}[1]{\mbox{\small \boldmath$#1$}}

\def\trace{\mbox{Tr}}

\usepackage{amscd,amsmath,amssymb}

\begin{document}

\title{Magnetothermal properties of the Heisenberg-Ising orthogonal-dimer
       chain with triangular $XXZ$ clusters}

\author{Vadim Ohanyan}
\affiliation{Department of Theoretical Physics, Yerevan State
University, Al. Manoogian 1, 0025, Yerevan, Armenia}
\email{ohanyan@yerphi.am}
\author{Andreas Honecker}
\affiliation{Institut f\"ur Theoretische Physik,
Georg-August-Universit\"at G\"ottingen, 37077 G\"ottingen, Germany}
\affiliation{Fakult\"at f\"ur Mathematik und Informatik,
             Georg-August-Universit\"at G\"ottingen,
             37073 G\"ottingen, Germany}
\date{March 21, 2012; revised July 22, 2012}

\begin{abstract}
We study a spin-1/2 model with triangular $XXZ$-clusters on the
orthogonal-dimer chain in the presence of an external magnetic field.
First, we discuss the case where the triangular clusters are coupled via
intermediate ``classical'' Ising spins. Diagonalization of the triangular
$XXZ$-clusters yields the exact ground states; finite-temperature
properties are computed exactly by an additional transfer-matrix step. A
detailed analysis reveals a large variety of ground states at
magnetization $M$ equal to fractions $0$, $1/4$, and $1/2$ of the
saturation magnetization $M=1$. Some of these ground states break
translational symmetry spontaneously and give rise to doubling of the unit
cell. In a second part we present complementary numerical data for the
spin-1/2 Heisenberg model on the orthogonal-dimer chain. We analyze
several examples of $T=0$ magnetization curves, entropy as a function of
temperature $T$ and magnetic field, and the associated magnetic cooling
rate. Comparison of the two models shows that in certain situations the
simplified exactly solvable model yields a qualitatively or sometimes even
quantitatively accurate description of the more challenging quantum model,
including a case which may be relevant to experimental observations of an
enhanced magnetocaloric effect in the two-dimensional compound
SrCu$_2$(BO$_3$)$_2$.
\end{abstract}

\pacs{75.10.Pq; 
75.30.Kz;   
75.40.Cx;   
75.40.Mg    
}

\maketitle

\section{Introduction}

There is a two-fold interest in the statistical mechanics of many-body
lattice spin systems. On the one hand, one-dimensional integrable
many-body models\cite{bax,tak} have been intensively investigated, if not
since the seminal 1931 paper of Hans Bethe,\cite{bethe} then at least
since the 1960-s. A linear chain of localized spins which interact with
each other via exchange interactions (Heisenberg chain) is the key
component of almost all such models. The research on integrable spin
chains is often considered to be separated from the underlying physical
context, as their integrability is deeply related to modern algebraic
structures (see for instance Refs.\ \onlinecite{isaev,QG}) and attracts
mainly mathematical physicists. On the other hand, various lattice spin
systems serve as the basis for \textit{physical models} for describing
magnetism of materials and related phenomena. As a rule, quantum lattice
spin models corresponding to real materials rarely admit exact solutions.
Therefore, one often has to resort to numerical calculations in order to
obtain reliable results for the thermodynamic properties of such strongly
correlated magnetic materials. However, for one-dimensional integrable
models, there are powerful mathematical tools allowing one, at least in
principle, to compute the thermodynamic functions exactly.\cite{tak, klu}
Despite their impressive efficiency, these methods are applicable only if
the underlying lattice model is integrable, which is not always the case
for real magnetic materials. However, there is another class of
application for so-called classical one-dimensional lattice models, namely
as effective descriptions of the conformations of
macromolecules.\cite{macro} In that case, one can obtain an analytical
expression for the partition function and, thus, for all thermodynamic
functions.

Lattice spin models with quantum and classical spins, \cite{DW75}
respectively Ising and Heisenberg \cite{oh03,ayd,str2, str, strMCE, roj,
ant09, oha09, bel09, roj2,chak12} or Hubbard interactions
\cite{per09,Lisnii11} combine properties of both quantum and classical
spin systems. They can be solved exactly by the classical transfer-matrix
method, while the entries of the transfer matrix contain the contribution
from the quantum clusters of the model. Such exact results obtained in
one-dimensional models containing both quantum and classical interactions
can shed light on the properties of the corresponding purely quantum ones
and as a result into the more profound understanding of the thermodynamic
behavior of magnetic materials with a one-dimensional arrangement of
exchange bonds.\cite{DW75,oh03, ayd, str2, str, strMCE, roj, ant09, oha09,
bel09, roj2,chak12,per09,Lisnii11} For instance, certain qualitative
properties of one-dimensional quantum spin systems can remain intact under
the replacement of some or even all quantum exchange interactions by Ising
ones. In particular, if the magnetization curve has an intermediate
plateau at some fixed value of the magnetization, the same feature may
persist also in the modified system, although the region of parameters at
which it appears can be different.\cite{oh03, str2} In some cases, one
even finds a rather good quantitative correspondence between the exact
magnetization curves obtained for the modified system and that obtained by
numerical methods for the underlying fully quantum system.\cite{str2}

Some real compounds may in fact be well approximated by such mixed
Heisenberg-Ising models. One example is given by a family of trimetallic
coordination polymer compounds.\cite{chem, ring, Dy_10}
Due to the presence of highly anisotropic Dy$^{3+}$ ions, these compounds
yield a direct realization of low-dimensional spin systems with Ising
and Heisenberg bonds. The single-chain magnet reported in Ref.\
\onlinecite{chem} is an example of a Heisenberg-Ising chain of triangles
with three-spin linear quantum clusters; another interesting
molecular magnet (spin ring) based on the highly anisotropic properties of the
Dy$^{3+}$ ions has been reported in Ref.\ \onlinecite{ring}. Both systems
have been analyzed by the transfer-matrix solution of
Heisenberg-Ising spin chains in Ref.\ \onlinecite{Dy_10}.
Another class of examples is given by chain magnets with two alternating
magnetic ions, thus giving rise to alternating quantum and classical
spins (see, e.g., Refs.~\onlinecite{RCRS04,JACS10,SSR11}).

\begin{figure}[tb!]
   \includegraphics[width=\columnwidth]{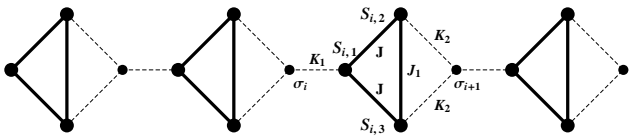}
\caption{``Orthogonal-dimer'' chain constructed from triangles of quantum
$s=1/2$ spins (large circles) interacting with each other via Heisenberg
$XXZ$ interactions (thick solid lines) and assembled to a chain by
``classical'' Ising spins (small circles). One coupling constant ($J_1$)
connects the quantum spins along a vertical line while a generally
different coupling constant ($J$) connects the other two bonds in the
quantum triangle. Further coupling constants are assigned to the Ising
bonds connecting left ($K_1$) and right ($K_2$) spins with the
corresponding ``classical'' spins $\sigma_i$ and $\sigma_{i+1}$.}
\label{fig1}
\end{figure}

Here we continue to apply the exact transfer-matrix calculation for
elucidating the thermodynamic and ground-state magnetic properties of
one-dimensional spin systems with Ising and Heisenberg bonds\cite{ant09,
oha09, bel09, roj2,chak12,Dy_10} and perform an analysis of the
similarities and discrepancies between magneto-thermal properties of the
exactly solvable Heisenberg-Ising model and the underlying purely quantum
Heisenberg system. We consider a one-dimensional $s=1/2$ chain consisting
of triangular clusters with $XXZ$ Heisenberg interactions and single
classical sites between them (see Fig.\ \ref{fig1}). Each spin at the
single site is connected by bonds of Ising type with the two spins of its
left triangle and with one spin of the right triangle. The corresponding
arrangement of sites into a chain is known as ``orthogonal-dimer''
chain.\cite{odc,odc2,odc3,HSR04} However, since the symmetry of the chain
is already reduced by the nature of the interactions, here we also allow
for more general interactions.

The magnetization curve of the antiferromagnetic quantum orthogonal-dimer
chain is famous for an infinite series of magnetization
plateaux.\cite{odc3} This infinite sequence of plateaux is caused by the
local conservation of the total spin on each vertical dimer: for each $k$,
a ground state of the infinite system can be constructed by a periodic
sequence of $k$ vertical dimers in the triplet state $s=1$ with two
singlet dimers at the end. In the Heisenberg-Ising case considered here,
only the main plateaux with magnetization equal to $0, 1/4$, and $1/2$ of
the quantum orthogonal-dimer chain are observed in the zero-temperature
magnetization curve. Depending on parameters of the models, we find
various ground states with 4 and 8 sites per unit cell realized at these
magnetization values. Some of these ground states give rise to a doubling
of the unit cell and a corresponding spatial modulation of the local
magnetization. The appearance of such structures may be attributed to the
left-right asymmetry for the triangular quantum clusters. Similar
observations were previously reported in Refs.\ \onlinecite{oha09,bel09}
for the sawtooth chain with Ising and Heisenberg bonds.

This paper is organized as follows. In section \ref{sec2} we present
the model and describe the exact computation of thermodynamic
properties using a classical transfer matrix. In section
\ref{sec:GS} we then specialize to zero temperature and describe the
ground states. Section \ref{sec4} contains a discussion of the
ground-state phase diagram in the presence of an external magnetic
field. Section \ref{sec5} presents  some exact magnetization curves
of the Heisenberg-Ising model for finite temperatures. Finally, in
section \ref{sec6}, we present numerical results for the full
quantum Heisenberg model and compare it to the Heisenberg-Ising
system. In this comparison, we focus on two properties: the
ground-state magnetization curve and the magnetocaloric effect. The
paper ends with a conclusion.

\section{The model and its exact solution}

\label{sec2}

We consider a spin chain consisting of $s=1/2$ $XXZ$-Heisenberg triangular
spin clusters, connected to each other via single spins situated between
the triangles in such a way that each left-hand single spin is connected
with the Ising bond $\sigma_i$-$S_i^z$ to the single spin of a triangular
cluster and each right-hand single spin is connected by the Ising bond to
two spins in the vertical line of the triangular cluster (see Fig.\
\ref{fig1}). This chain is topologically equivalent to the famous
orthogonal-dimer chain,\cite{odc,odc2,odc3,HSR04} albeit with more general
interactions. The Hamiltonian of the system is conveniently represented as
a sum of the single-block Hamiltonians $\mathcal{H}_i$
\bea
 \mathcal{H}=\sum_{i=1}^L\left(
{\mathcal{H}}_i-\frac{H_1}{2}\left(\sigma_i + \sigma_{i+1}
\right)\right) \, ,
\label{H}
\eea
where
\bea
{\mathcal{H}}_i&=&J\,\left\{\Delta \left[ S_{i,1}^x (
S_{i,2}^x+S_{i,3}^x)+S_{i,1}^y (
S_{i,2}^y+S_{i,3}^y) \right] \right. \nonumber \\
&& \left. +S_{i,1}^z(S_{i,2}^z+S_{i,3}^z) \right\} \nonumber \\
&& +
J_1\,\left(\Delta_1(S_{i,2}^xS_{i,3}^x+S_{i,2}^yS_{i,3}^y)+S_{i,2}^zS_{i,3}^z\right)
\nonumber \\
&& + K_1\,\sigma_i^zS_{i,1}^z+K_2\,\sigma_{i+1}^z
(S_{i,2}^z+S_{i,3}^z) \nonumber \\
&&
-H_2 (S_{i,1}^z+S_{i,2}^z+S_{i,3}^z) \, .
\label{Hi}
\eea
Here $L$ is the number of blocks (cells) in the chain (the number of spins
$N$ is $4L$), $\bmth{S}_{i,a}$ stands for the spin
operators of the $i$-th triangle in the chain, $a=1,2,3$ numbers the
three sites of the triangle, and $\sigma_i^z$ is the $z$-component
of the single spin situated at the left-hand side of the $i$-th
triangle. Here, we allow the exchange
interaction constants between the spins on the vertical line $J_1$,
$\Delta_1$ and between the single spin of a triangle and spins on the
vertical line $J$, $\Delta$ to be different as well as the interaction
constants for left and right single spins, $K_1$ and $K_2$,
respectively. Magnetic fields acting on $S$-spins and $\sigma$-spins
are also considered to be different.

We are going to derive exact expressions for all thermodynamic functions
of the chain. To this end one should calculate the partition function
\bea
\mathcal{Z}=\sum_{(\sigma)} \trace_{\bmths{S}}
 e^{-\beta \sum_{i=1}^L\left( {\mathcal{H}}_i-\frac{H_1}{2}\left(\sigma_i + \sigma_{i+1} \right)\right)} \, . \label{Z}
\eea
Spin variables of single spins can be regarded as classical
Ising ones, as they contribute only their diagonal $z$-component
and hence can be replaced with their
eigenvalues $\pm 1/2$. Thus, in the partition function one should
sum over all values of the $\sigma_i$ and take a trace over the
states of all spin operators $\bmth{S}$. Due to the special
structure of interactions,
the partition function can be evaluated exactly. To proceed, we first note
that the Hamiltonians of different blocks commute
\bea
[\mathcal{H}_i, \mathcal{H}_j]=0 \, . \label{com}
\eea
Thus, one can
expand the exponential in Eq.\ (\ref{Z}) and get the following
expression
\bea \mathcal{Z}=\sum_{(\sigma)}\prod_{i=1}^L e^{\beta
\frac{H_1}{2}\left(\sigma_i+\sigma_{i+1} \right)}\trace_i e^{-\beta
\mathcal{H}_i}, \label{Ze}
\eea
where $\trace_i$ implies the trace
over all states of the $i$-th block of the chain. The trace can
be easily calculated once the block Hamiltonian
$\mathcal{H}_i$ is diagonalized, which can be achieved analytically
in our case. One finds the following eigenvalues
\bea
\lambda_{1,2}\left(\sigma_i, \sigma_{i+1} \right)&=&\frac{1}{4}\left( J_1+2J\right)
\nonumber \\
&& \pm\frac{1}{2}\left(K_1\sigma_i+2 K_2 \sigma_{i+1}\right)
\mp\frac{3}{2}H_2, \nonumber \\
\lambda_{3,4}\left(\sigma_i, \sigma_{i+1} \right)&=&-\frac{1}{4}J_1\left(1+2\Delta_1 \right)
 \pm\frac{1}{2}K_1\sigma_i\mp\frac{1}{2}H_2,\nonumber \\
\lambda_{5,6}\left(\sigma_i, \sigma_{i+1} \right)&=&\frac{1}{4}\left(J_1\Delta_1-J-Q_{\mp \sigma_i, \pm \sigma_{i+1}}
\right)
\nonumber \\
&& \pm\frac{1}{2}K_2\sigma_{i+1}
\mp\frac{1}{2}H_2, \nonumber \\
\lambda_{7,8}\left(\sigma_i, \sigma_{i+1}\right)
&=&\frac{1}{4}\left(J_1\Delta_1-J+Q_{\mp \sigma_i,
\pm \sigma_{i+1}} \right) \nonumber \\
 &&\pm\frac{1}{2}K_2\sigma_{i+1}
 \mp\frac{1}{2}H_2, 
 \label{eval} \eea
where the following abbreviation is adopted:
\begin{equation}
 Q_{\sigma,\sigma^\prime}=\sqrt{8 J^2 \Delta^2+(J_1 (1-\Delta_1)-J+2(K_1\sigma+K_2\sigma^\prime))^2}.
\label{Q}
\end{equation}
The corresponding eigenstates carry a definite
value of $S_{\text{tot}}^z=S_1^z+S_2^z+S_3^z$:
the eigenvalues
$\lambda_{1,2}$ correspond to states with
$S_{\text{tot}}^z=\pm 3/2$  and the remaining eigenvalues
to $S_{\text{tot}}^z=\pm 1/2$. The eigenvectors read:
\begin{equation}
\begin{array}{rrl}
\lambda_1:\     & |v_{\frac{3}{2}}\rangle&=\displaystyle |\uparrow\uparrow\uparrow\rangle,    \\
\lambda_2:\     & |v_{-\frac{3}{2}}\rangle&=\displaystyle |\downarrow\downarrow\downarrow\rangle,   \\
\lambda_3:\     & |v_{\frac{1}{2},a}\rangle&=\displaystyle \frac{1}{\sqrt{2}}\left(|\uparrow\uparrow\downarrow\rangle-|\uparrow\downarrow\uparrow\rangle \right),  \\
\lambda_4:\     & |v_{-\frac{1}{2},a}\rangle&=\displaystyle \frac{1}{\sqrt{2}}\left(|\downarrow\downarrow\uparrow\rangle-|\downarrow\uparrow\downarrow\rangle \right),   \\
\lambda_{5,7}:\ & |v_{\frac{1}{2},s}^{\pm}\rangle&=\displaystyle \frac{1}{\sqrt{2+c_{\pm}^2}}\left(|\uparrow\uparrow\downarrow\rangle+|\uparrow\downarrow\uparrow\rangle
+c_{\pm}|\downarrow\uparrow\uparrow\rangle \right),  \\
\lambda_{6,8}:\ & |v_{-\frac{1}{2},s}^{\pm}\rangle&=\displaystyle \frac{1}{\sqrt{2+\overline{c}_{\pm}^2}}\left(|\downarrow\downarrow\uparrow\rangle+|\downarrow\uparrow\downarrow\rangle
+\overline{c}_{\pm}|\uparrow\downarrow\downarrow\rangle\right) \, .
\end{array}
\label{evec}
\end{equation}
In each state $|\cdots\rangle$ the first, second, and
third arrow denotes an $z$-eigenbasis of $\bmth{S}_{i,1}$,
$\bmth{S}_{i,2}$, and $\bmth{S}_{i,3}$,
respectively. Furthermore, we have introduced the following
abbreviations
\bea
c_{\pm}&=&\frac{J_1(1-\Delta_1)-J+2(K_2\sigma_{R}-K_1\sigma_L)\pm
Q_{-\sigma_L,\sigma_R}}{J
\Delta}, \nonumber \\
\overline{c}_{\pm}&=&\frac{J_1(1-\Delta_1)-J-2(K_2\sigma_{R}-K_1\sigma_L)\pm
Q_{\sigma_L,-\sigma_R}}{J \Delta} \, . \nonumber
\eea
Here $\sigma_{R(L)}$ stands for the value of the right (left)
$\sigma$-spin for the given triangle.

After calculating the trace over the quantum degrees of freedom one
arrives at the following expression for the partition function \bea
\mathcal{Z}=\sum_{(\sigma)}\prod_{i=1}^L e^{\beta
\frac{H_1}{2}\left(\sigma_i+\sigma_{i+1} \right)} Z_{\sigma_i,
\sigma_{i+1}}= \trace \bmth{T}^L, \label{Zf} \eea
 where ${\bmth{T}}$
is a $2\times2$ transfer matrix and $Z_{\sigma_i, \sigma_{i+1}}$ is
the ``partial'' partition function of the $i$-th block: \bea
Z_{\sigma_i, \sigma_{i+1}} =\trace_i e^{-\beta
\mathcal{H}_i}=\sum_{n=1}^8 e^{-\beta \lambda_n \left( \sigma_i,
\sigma_{i+1} \right)} \, . \label{Sp} \eea Substituting $\lambda_n
\left(\sigma_i, \sigma_{i+1} \right)$ from Eq.\ (\ref{eval}) one
gets
\begin{widetext}
\bea Z_{\sigma_i, \sigma_{i+1}}
&=&2\left\{e^{-\beta\frac{J_1+2J}{4}}\cosh\left(\beta\frac{3H_2-K_1
\sigma_i-2K_2 \sigma_{i+1}}{2}\right)+e^{\beta\frac{J_1
(1+2\Delta_1)}{4}}\cosh\left(\frac{H_2-K_1\sigma_i}{2} \right)
\right. \\ \nonumber &&\left.
+e^{-\beta\frac{J_1\Delta_1-J}{4}}\left[e^{\beta \frac{H_2+K_2
\sigma_{i+1}}{2}}\cosh \left( \beta \frac{Q_{-
\sigma_i,\sigma_{i+1}}}{4}\right) +e^{-\beta \frac{H_2+K_2
\sigma_{i+1}}{2}}\cosh \left( \beta \frac{Q_{\sigma_i,
-\sigma_{i+1}}}{4}\right)\right]\right\} \, . \label{AZ} \eea
\end{widetext}
According to  Eq.\ (\ref{Zf}), the
partial partition functions for one block multiplied by the factor
$e^{\beta \frac{H_1}{2}\left(\sigma_i+\sigma_{i+1} \right)}$
yield the entries of the transfer matrix for a certain block:
\begin{equation}
{\bmth{T}}=\left( \begin{array}{lcr}
      e^{\beta  \frac{H_1}{2}}Z_{\frac{1}{2},\frac{1}{2}}  & Z_{\frac{1}{2}, -\frac{1}{2}} \\
      Z_{-\frac{1}{2},\frac{1}{2}}  & e^{-\beta  \frac{H_1}{2}}Z_{-\frac{1}{2},-\frac{1}{2}} \label{T}
      \end{array}
\right) \, .
\end{equation}
Thus, calculation of the partition function for the
system under consideration boils down to evaluating the
trace of the $L$-th power of a 2 by 2 transfer matrix in full
analogy with the ordinary Ising chain (see for example Refs.\
\onlinecite{bax,Huang}). The free energy per
unit cell in the thermodynamic limit is obtained from
the maximal eigenvalue of the transfer matrix (\ref{T}):
\begin{widetext}
\begin{equation}
 f=-\frac{1}{\beta}\ln\left[\frac{1}{2}\left(e^{\beta \frac{H_1}{2}} {Z_{\frac{1}{2},\frac{1}{2}}}
  + e^{-\beta \frac{H_1}{2}} {Z_{-\frac{1}{2},-\frac{1}{2}}}
 +\sqrt{\left(e^{\beta \frac{H_1}{2}} {Z_{\frac{1}{2},\frac{1}{2}}}
  - e^{-\beta \frac{H_1}{2}} {Z_{-\frac{1}{2},-\frac{1}{2}}} \right)^2
  +4 \, {Z_{\frac{1}{2}, -\frac{1}{2}}}\, {Z_{-\frac{1}{2},\frac{1}{2}}}}
     \right)\right] \, . \label{f}
\end{equation}
\end{widetext}
Now it is straightforward to obtain not only all thermodynamic properties,
but also expressions for the sublattice and total magnetizations:
\bea
M_{\sigma}&=&\frac{2}{L\mathcal{Z}}\sum_{(\sigma)}\trace_{\bmths{S}}\left(\sum_{i=1}^{L}\sigma_{i}e^{-\beta\mathcal{H}}\right)=
 -2\left(\frac{\partial f}{\partial H_{1}}\right)_{\beta,H_{2}}, \nonumber \\
M_S&=&\frac{2}{3L\mathcal{Z}}\sum_{(\sigma)}\trace_{\bmths{S}}\left(\sum_{i=1}^{L}\sum_{a=1}^3(S_{i,a}^z)e^{-\beta\mathcal{H}}\right)\nonumber
 \\
&=& -\frac{2}{3}\left(\frac{\partial f}{\partial H_{2}}\right)_{\beta,H_{1}},  \nonumber \\
M&=&\frac{1}{2L\mathcal{Z}}\sum_{(\sigma)}\trace_{\bmths{S}}\left(\sum_{i=1}^{L}
\left(\sigma_i+\sum_{a=1}^3(S_{i,a}^z)\right)e^{-\beta\mathcal{H}}\right)\nonumber
\\
&=&\frac{1}{4}M_{\sigma}+\frac{3}{4}M_S\,.
\label{MMM}
\eea

\section{Ground-state properties}

\label{sec:GS}

\begin{table*}[p]
\caption{Ground states of the system and their energies per
block.\cite{footnote1} The function $Q$ is defined in Eq.~(\ref{Q}).
Vertical ellipses denote the quantum singlet state of the
corresponding spins; horizontal rectangles denote a disordered pair
of two Ising spins with $S_{\text{tot}}=0$; gray triangles stand for
the $|v_{\pm\frac{1}{2},s}^{-}\rangle$ ground states and a frame
around a triangle indicates the degenerate superposition of
$|v_{\frac{1}{2},s}^{-}\rangle$ and $|v_{\frac{1}{2},a}^{-}\rangle$.
For more details compare the text.} \centering
\begin{ruledtabular}
\begin{tabular}{c c c c}
\hline
Label &   Energy per block & Spatial period & Figure \\ [0.5ex]
\hline
AF1 & $-\frac{1}{4}\left(J_1(1+2 \Delta_1)+K_1 \right)$ & undefined &
\lower 1.6pc\hbox{\includegraphics[width=7cm]{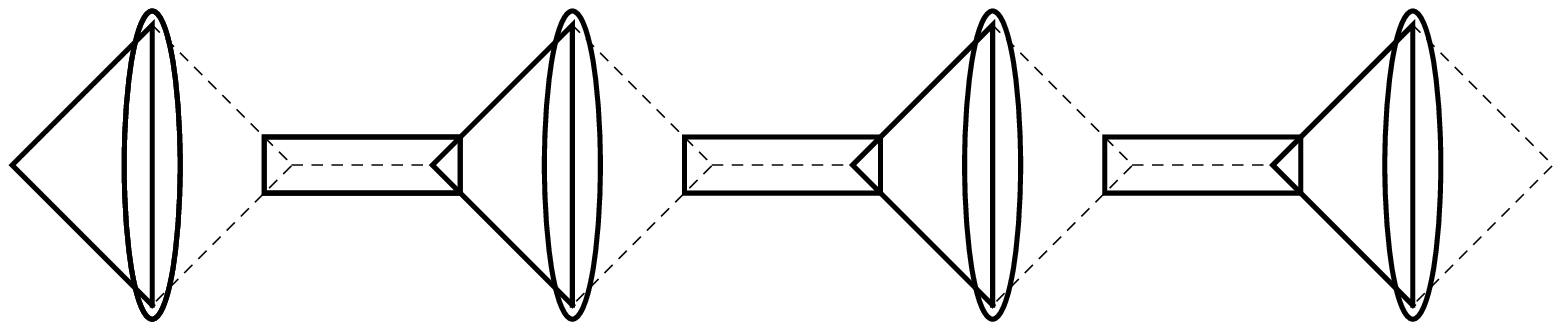}}  \\
\hline
 AF2 & $-\frac{1}{4}\left(J-J_1\Delta_1
  +K_2+ Q_{1/2,-1/2}\right)$ & 1 & \lower 1.4pc\hbox{\includegraphics[width=7cm]{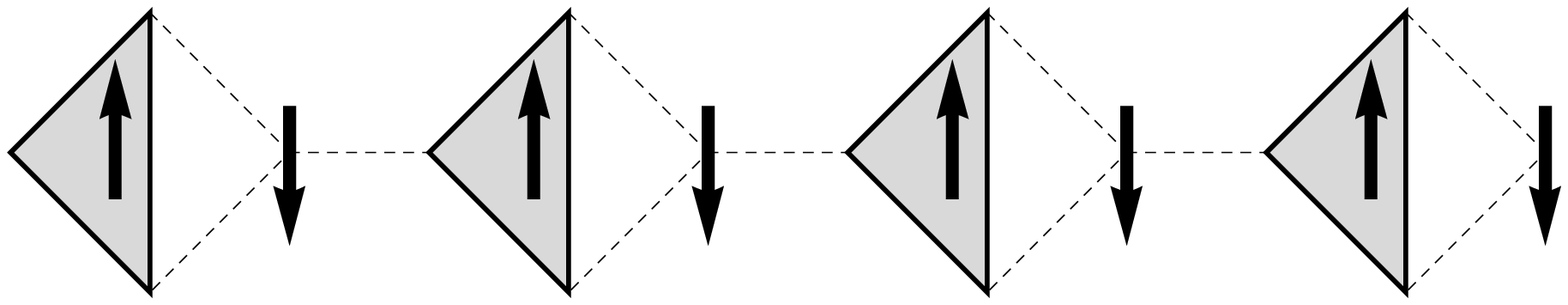}}  \\
\hline AF3 & $-\frac{1}{4}\left(J-J_1\Delta_1
  +K_2+Q_{-1/2,-1/2} \right)$ & 2  & \lower 1.4pc\hbox{\includegraphics[width=7cm]{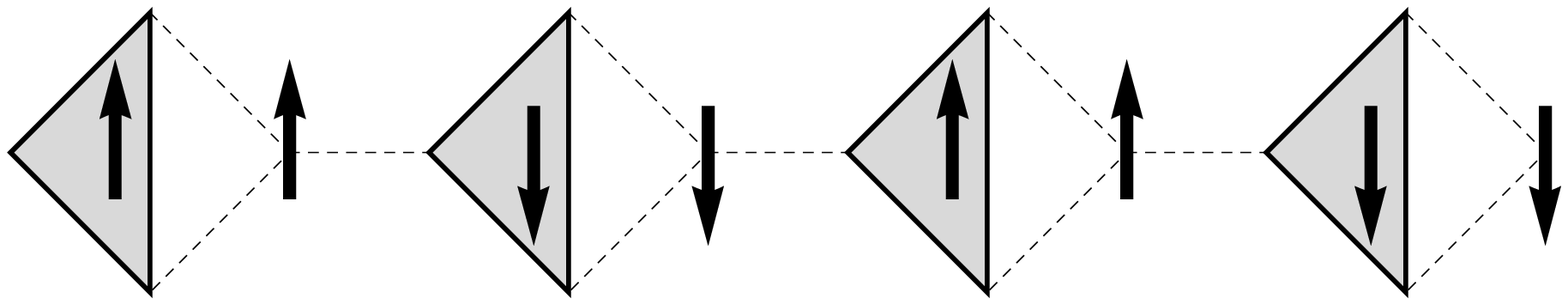}} \\
\hline AF4 & $-\frac{1}{4}\left(J-J_1\Delta_1
  -K_2+Q_{1/2,1/2} \right)$ & 2 &  \lower 1.4pc\hbox{\includegraphics[width=7cm]{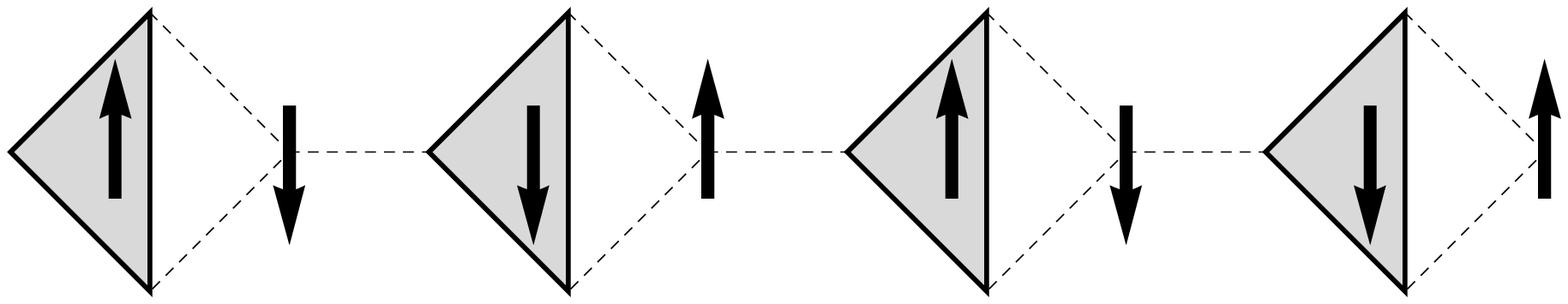}} \\
\hline AF5 & $\frac{1}{4}\left(J_1+2 J-K_1+2
K_2\right)$ & 2 &  \lower 1.8pc\hbox{\includegraphics[width=7cm]{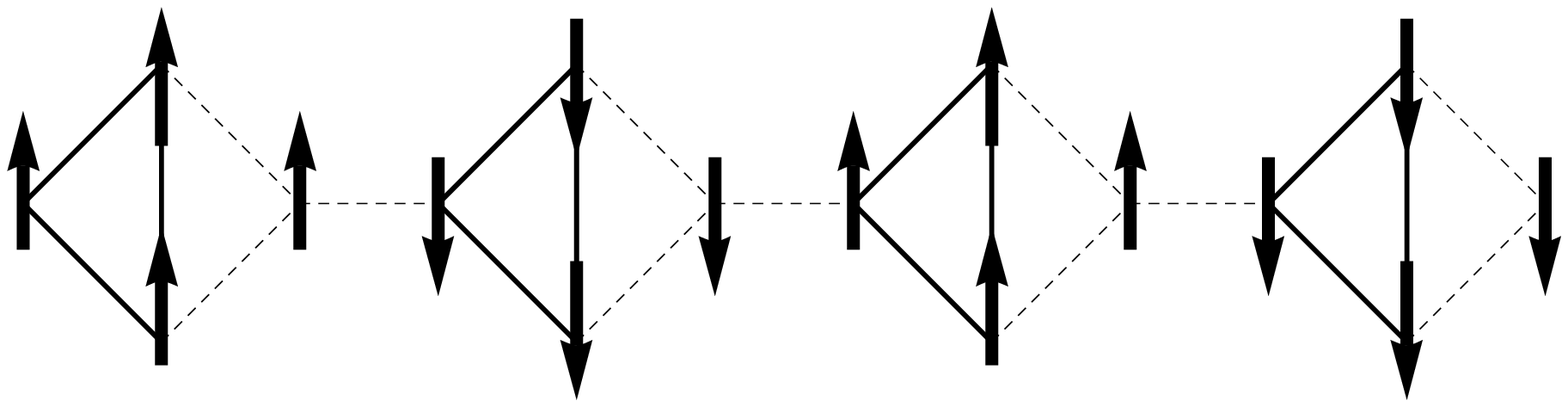}} \\
\hline
SM1 & $-\frac{1}{4}\left(J-J_1\Delta_1+Q_{1/2,1/2}+Q_{-1/2,-1/2} \right)-\frac{1}{2}H$ & 2 &  \lower 1.4pc\hbox{\includegraphics[width=7cm]{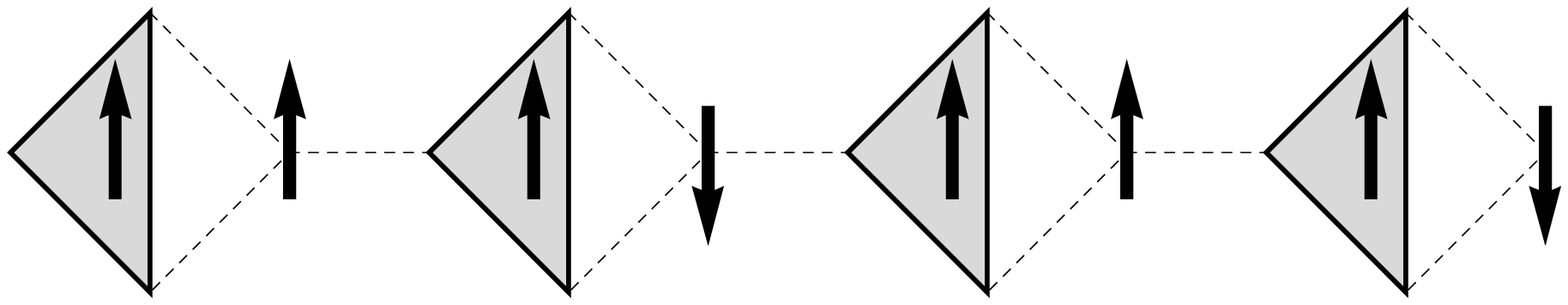}} \\
\hline
SM2 & $-\frac{1}{8}\left(J+J_1\left(1+\Delta_1 \right)+K_1+K_2+Q_{-1/2,-1/2}\right)-\frac{1}{2}H$ & 2 &  \lower 1.4pc\hbox{\includegraphics[width=7cm]{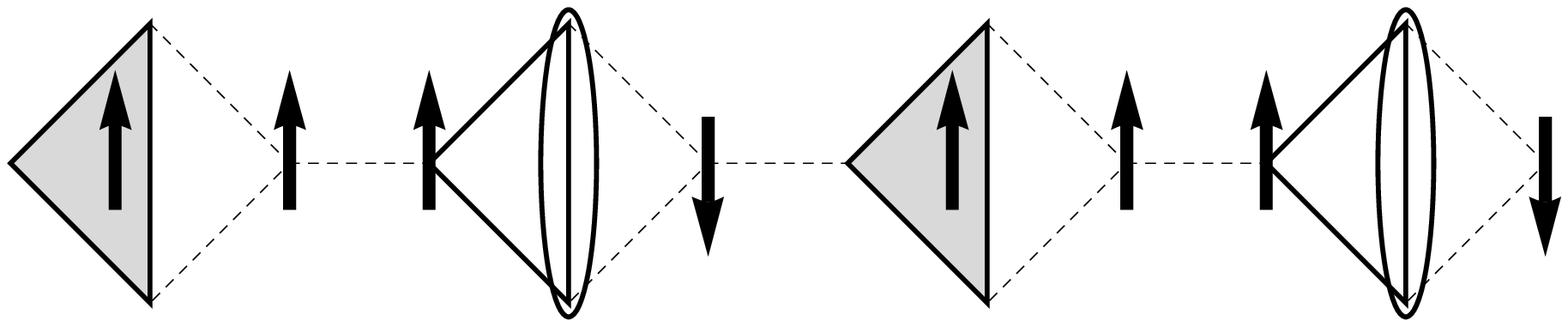}} \\
\hline
SM3 & $\frac{1}{4}\left(J-J_1\Delta_1-K_1+K_2\right)-\frac{1}{2}H$ & 2 &  \lower 1.6pc\hbox{\includegraphics[width=7cm]{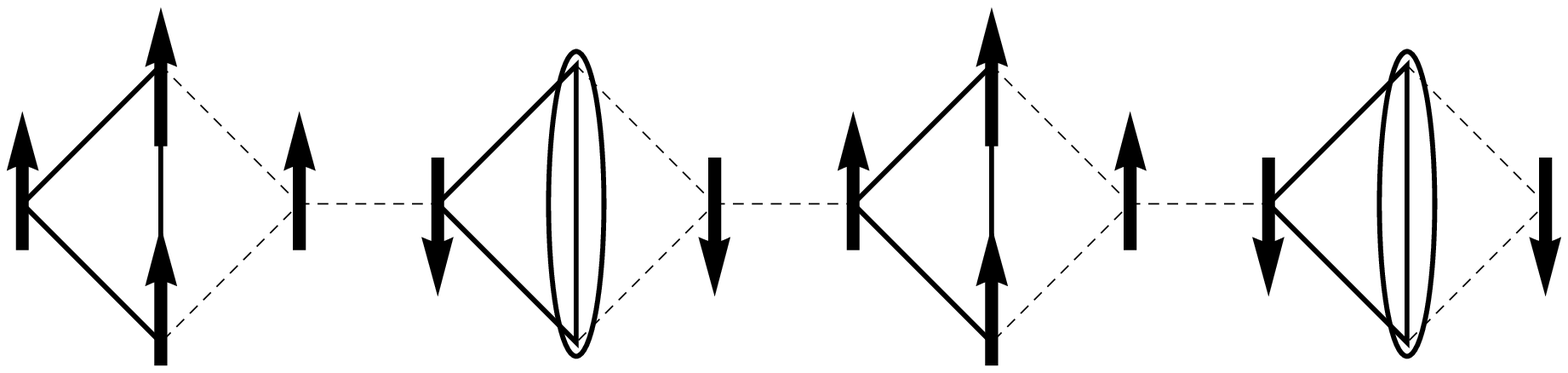}} \\
\hline
F1  & $\frac{1}{4}\left(J_1+2J-K_1-2K_2 \right)-H$ & 1 &   \lower 1.6pc\hbox{\includegraphics[width=7cm]{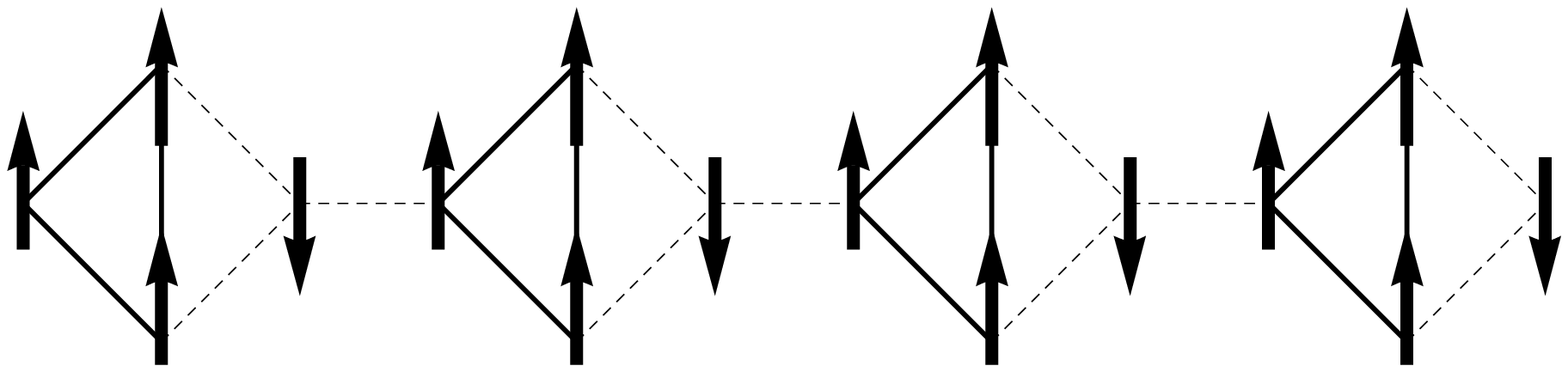}}  \\
\hline
F2  & $-\frac{1}{4}\left(J_1\left(1+2\Delta_1 \right)-K_1 \right)-H$ & 1 &  \lower 1.6pc\hbox{\includegraphics[width=7cm]{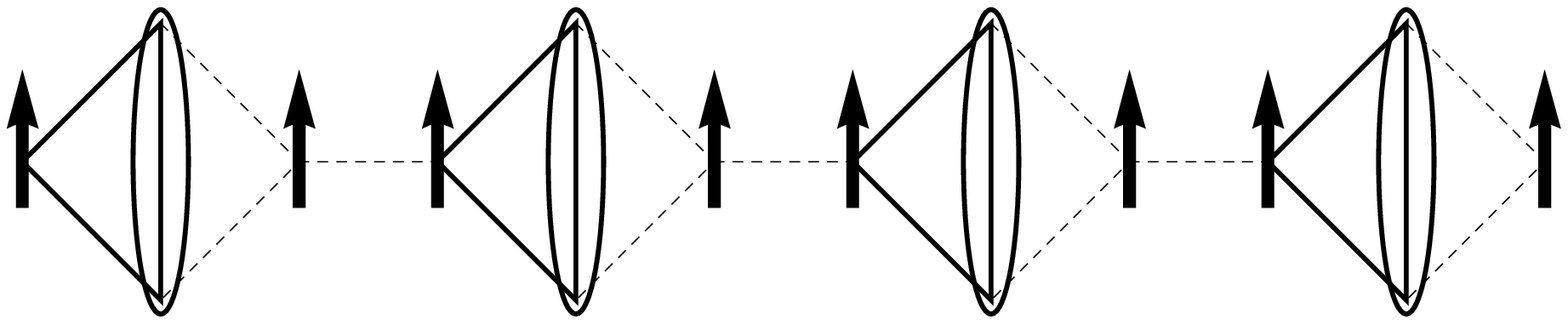}} \\
\hline
F3  & $-\frac{1}{4}\left(J-J_1\Delta_1-K_2+Q_{-1/2,1/2} \right)-H$ & 1 &  \lower 1.4pc\hbox{\includegraphics[width=7cm]{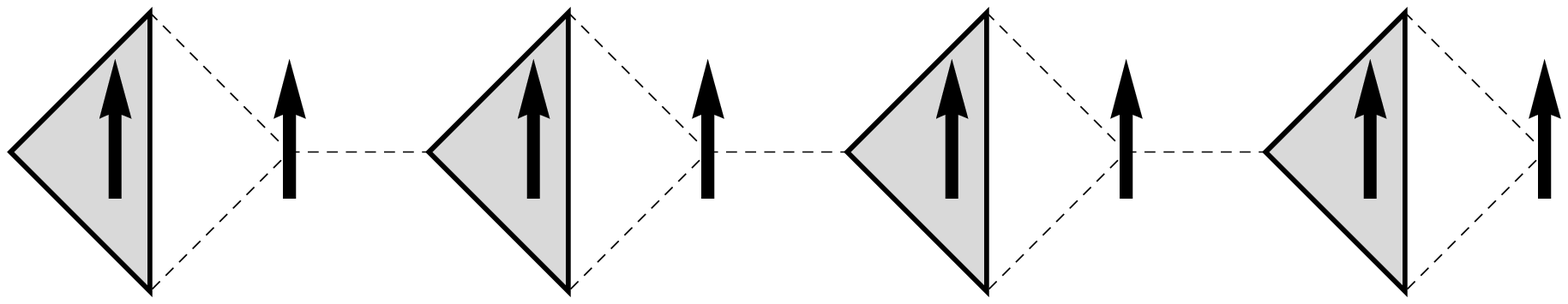}} \\
\hline F4  & $-\frac{1}{4}\left(J\left(1+2\Delta \right)-K
\right)-H$ & undefined & \lower 1.8pc\hbox{\includegraphics[width=7cm]{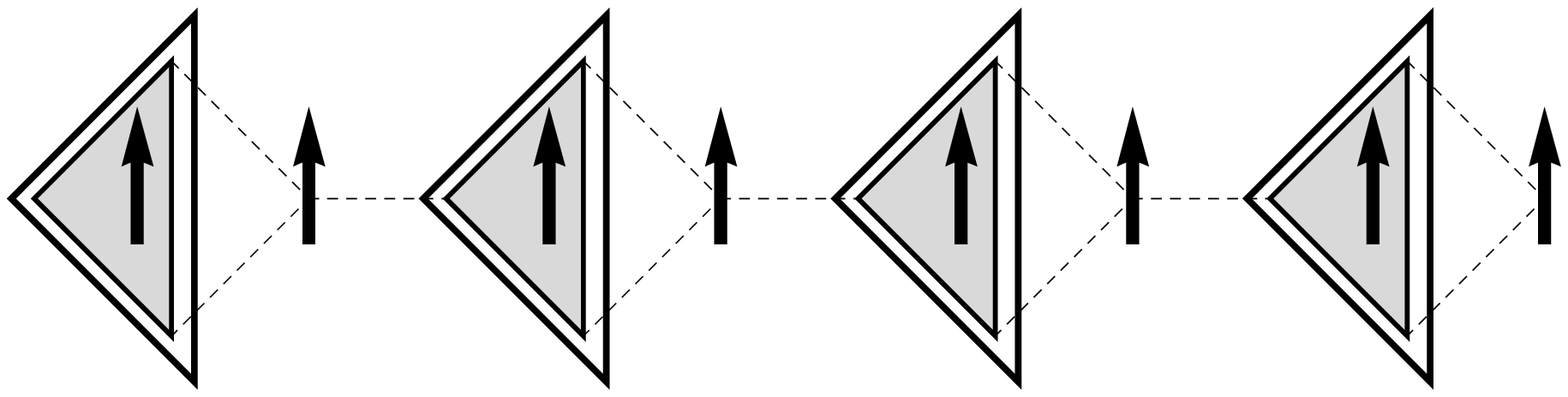}} \\
\hline
SP  & $\frac{1}{4}\left(J_1+2J+K_1+2K_2 \right)-2H$ & 1 & \lower 1.6pc\hbox{\includegraphics[width=7cm]{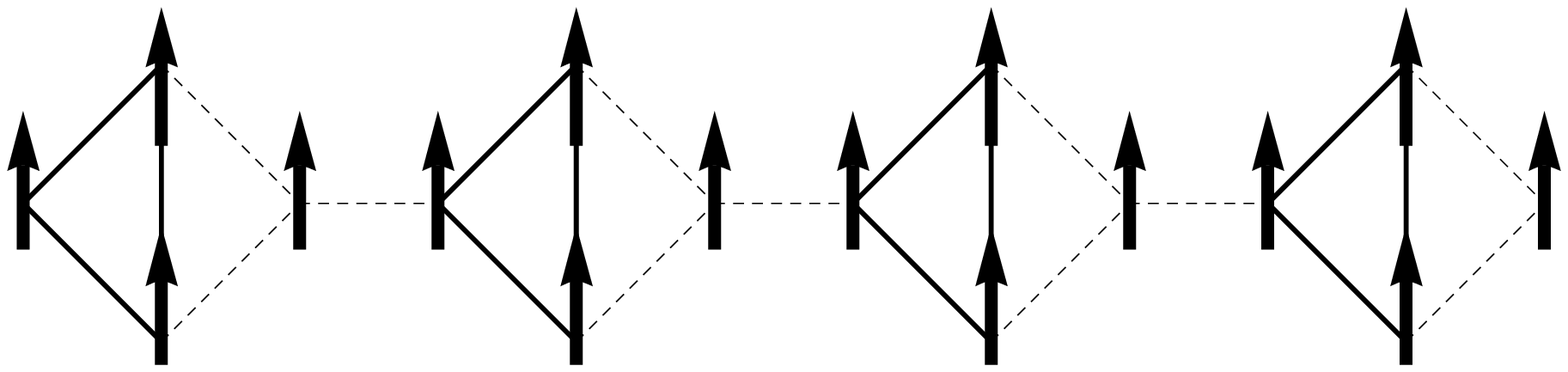}} \\[1ex]
\hline
\end{tabular}
\end{ruledtabular}
\label{table}
\end{table*}

{}From the eigenvectors for one block of the system one can
construct all possible ground states of the whole chain. The ground
states of the system under consideration can be classified by the
values of magnetization, $M$, which we normalize to saturation values $\pm 1$.
In particular, one finds antiferromagnetic $M=0$, ferrimagnetic ($M=1/2$ or $1/4$),
and ferromagnetic $M=1$ ground states.

There are five ``antiferromagnetic'' ground states.
Among them, there is a special one
which is the closest analog of the fully dimerized ground state of
the purely quantum orthogonal-dimer chain.\cite{odc,odc2,odc3}
However, in contrast to the latter, here the corresponding ground
state exhibits a two-fold degeneracy per block due to the Ising-type
interactions on the horizontal dimers. So, we will refer to this
ground state as the {\em degenerate antiferromagnetic} state, which we will abbreviate by
$|\text{AF1}\rangle$. In the state $|\text{AF1}\rangle$,
the spins on vertical dimers form a perfect
spin singlet, while the spins connected by a horizontal Ising bond are
aligned antiparallel,  but can
freely change their overall orientation without changing the
energy of the ground state. The corresponding wave functions read:
\begin{equation}
 |\text{AF1}\rangle=\prod_{i=1}^L|v_{\pm\frac{1}{2},a},\mp\rangle_i
 \, . \label{AF1}
\end{equation}
Here, $\mp$ stands for the ``down'' (``up'') orientation for the
$\sigma$ spin from the $i$-th block, which appears simultaneously
with the $|v_{\frac{1}{2},a}\rangle_i$
$(|v_{-\frac{1}{2},a}\rangle_i)$ configuration of the $XXZ$-triangle
of the same block (compare Eq.~(\ref{evec}) for the explicit form of
$|v_{\pm\frac{1}{2},a}\rangle_i$). It is worth mentioning, however,
that the disentangled nature of the antiferromagnetic state of two
spins connected by the Ising bond limits the
analogy between the AF1 and the perfectly dimerized ground state of the purely
quantum orthogonal-dimer chain. In particular, the two-fold
degeneracy per block in our case has its origin in the above
mentioned difference. Schematic figures illustrating the spin
arrangement for the various ground states as well as their energies
per block\cite{footnote1} can be found in Table \ref{table}.

There is another $M=0$
ground state with spatial period equal to the period of the system
(four spins in the unit cell). This ground state corresponds to
$S_{\text{tot}}^z=1/2$ states of the quantum triangles which are aligned
antiparallel with the $s=1/2$ $\sigma$-spins between them:
\begin{equation}
 |\text{AF2}\rangle=\prod_{i=1}^L|v_{\frac{1}{2},s}^-,\downarrow\rangle_i \, .
\end{equation}

Three further antiferromagnetic ground states are based on
the eigenvectors of the triangular clusters
which are asymmetric with respect to left and right $\sigma$-spins.
These ground states break the translational symmetry  spontaneously,
leading to a doubling of the unit cell
\bea
 |\text{AF3}\rangle&=&\prod_{i=1}^{L/2}|v_{\frac{1}{2},s}^-,\uparrow\rangle_{2i-1}|v_{-\frac{1}{2},s}^-,\downarrow\rangle_{2i},\\
 |\text{AF4}\rangle&=&\prod_{i=1}^{L/2}|v_{\frac{1}{2},s}^-,\downarrow\rangle_{2i-1}|v_{-\frac{1}{2},s}^-,\uparrow\rangle_{2i},
  \nonumber \\
 |\text{AF5}\rangle&=&\prod_{i=1}^{L/2}|v_{\frac{3}{2}},\uparrow\rangle_{2i-1}|v_{-\frac{3}{2}},\downarrow\rangle_{2i}\, .\nonumber
\label{AF34}
\eea
The $|\text{AF3}\rangle$ ground state is characterized by an
antiferromagnetically ordered sublattice of triangles with
$S_{\text{tot}}^z=\pm 1/2$ as well as by the same ordering of
$\sigma$-spins. These two sublattices form an ``up-up-down-down'' overall
arrangement of the local $s=1/2$ magnetic moments. One can also notice
that here we have an antiferromagnetic ordering of blocks with overall
spin equal to $\pm 1$. In the $|\text{AF4}\rangle$ ground state, as in the
previous case, one can see separate antiferromagnetic ordering of both
$\sigma$-spins and triangles which are in the $s=1/2$ state. However, inside
each block they compensate each other (see Table \ref{table}). In the
$|\text{AF5}\rangle$ ground state one can see antiferromagnetic ordering
of fully polarized blocks.

Now we turn to $M=1/4$. Since this yields a fractional spin polarization per unit cell,
$M=1/4$ ground states are accompanied by spontaneous breaking of translational symmetry,\cite{oya,HSR04}
more precisely, doubling of the unit cell. In view of the spatial modulation of the local spin polarization,
we will refer to the corresponding ground states as spin modulated ones:
\bea
 |\text{SM1}\rangle&=&\prod_{i=1}^{L/2}|v_{\frac{1}{2},s}^-,\uparrow\rangle_{2i-1}|v_{\frac{1}{2},s}^-,\downarrow\rangle_{2i},
 \label{eq:SM} \\
 |\text{SM2}\rangle&=&\prod_{i=1}^{L/2}|v_{\frac{1}{2},s}^-,\uparrow\rangle_{2i-1}|v_{\frac{1}{2},a},\downarrow\rangle_{2i},\nonumber \\
 |\text{SM3}\rangle&=&\prod_{i=1}^{L/2}|v_{\frac{3}{2}},\uparrow\rangle_{2i-1}|v_{-\frac{1}{2},a}^-,\downarrow\rangle_{2i}.\nonumber
\eea
These ground states have more complicated local spin arrangements than the antiferromagnetic ones. In contrast to the latter,
these ``spin-modulated'' ground states posses a doubled unit cell in which two blocks are not connected to each other by spin inversion
(see Table \ref{table}).
Our model exhibits also three translationally invariant ferrimagnetic ground states with $M=1/2$:
\bea
 |\text{F1}\rangle &=&\prod_{i=1}^L |v_{\frac{3}{2}},\downarrow\rangle_i,
 \label{eq:F} \\
 |\text{F2}\rangle &=&\prod_{i=1}^L |v_{\frac{1}{2},a},\uparrow\rangle_i, \nonumber \\
 |\text{F3}\rangle &=&\prod_{i=1}^L |v_{\frac{1}{2},s}^-,\uparrow\rangle_i \, . \nonumber
\eea
Furthermore, there is also one family of ground states
with a two-fold degeneracy per block
with $M=1/2$ in which all $\sigma$ spins point up,
while the quantum triangles can freely and independently  oscillate between $|v_{\frac{1}{2},a}\rangle$ and $|v_{\frac{1}{2},s}^-\rangle$.
However, these ground states appear only for the highly symmetric choice of parameters
$\Delta=\Delta_1$, $J=J_1$, $K_1=K_2$. The corresponding eigenvectors are
\begin{equation}
|\text{F4}\rangle =\prod_{i=1}^L
|V_{\frac{1}{2}},\uparrow\rangle_i,
\label{eq:F4}
\end{equation}
where $|V_{\frac{1}{2}}\rangle$ can be either one of the states
$|v_{\frac{1}{2},a}\rangle$ or $|v_{\frac{1}{2},s}^-\rangle$.
Finally, there is the fully polarized ground state
\begin{equation}
|\text{SP}\rangle=\prod_{i=1}^L |v_{\frac{3}{2}},\uparrow\rangle_i \, .
\end{equation}

\begin{figure}[tb!]
\begin{center}
\includegraphics[width=0.9\columnwidth]{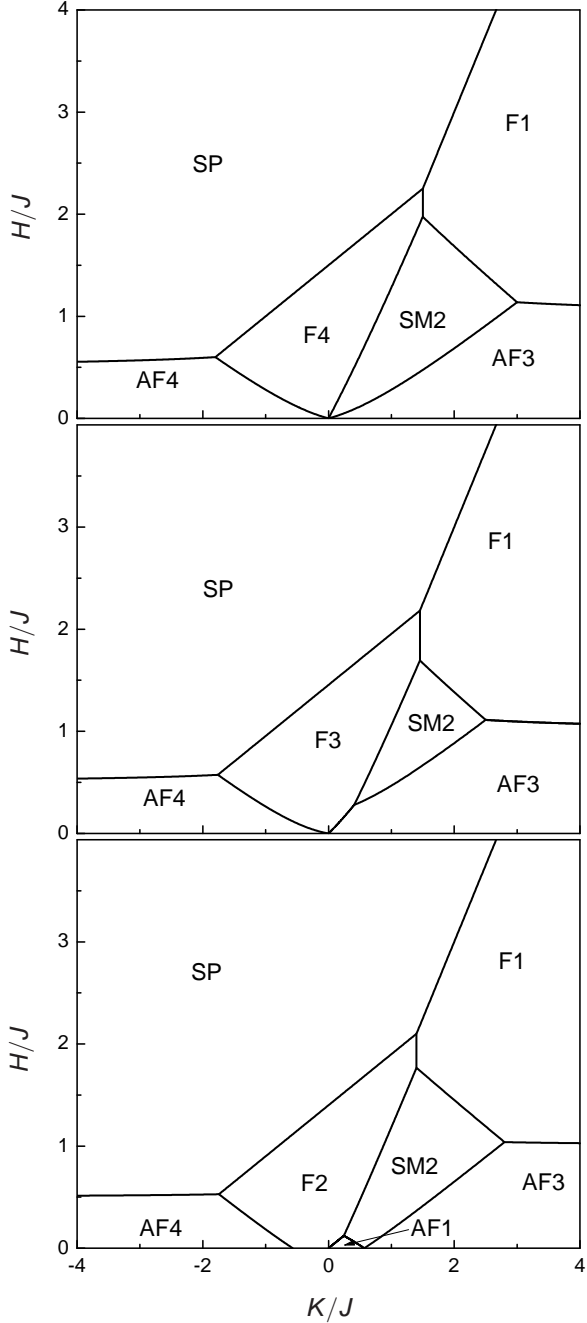}
\end{center}
\caption{\label{Fig-2} Ground-state phase diagrams for
antiferromagnetic $J_1=J>0$ and $K_1=K_2 \equiv K$. The top
panel shows the case of isotropic interaction ($\Delta=\Delta_1=1$).
A characteristic feature is the appearance of the
macroscopically degenerate ferrimagnetic phase F4 with $M=1/2$. The
lower two panels show the case of Ising-like anisotropies,
namely $\Delta=0.8 > \Delta_1=0.5$ (middle panel) and
$\Delta=0.5 < \Delta_1=0.8$  (lowest panel).
Note that $\Delta$
and $\Delta_1$ are the prefactors of the $XY$-term in the spin
Hamiltonian.
}
\end{figure}

\section{Ground-state phase diagrams}

\label{sec4}

Ground-state phase diagrams can be derived from the energies
of the ground states given in  Table \ref{table}.
Since there are many free parameters, one needs to make some choices
and it is natural to set certain exchange constants equal.
We are going to focus on two cases. The first one corresponds to a situation with one
exchange constant for all quantum bonds and another one for all Ising bonds
($J=J_1\equiv J$, $K_1=K_2\equiv K$)
while the anisotropy constants for the two types of quantum bonds can still be
different. The other case is well investigated in the context of the quantum
orthogonal-dimer chain,\cite{odc,odc2,odc3,HSR04}
where there are two interaction parameters, one for vertical and horizontal dimers,
and another one for diagonal bonds ($K_1=J_1\equiv \tilde{J}$, $K_2=J\equiv K$).

\subsection{$J=J_1$, $K_1=K_2 \equiv K$}

\begin{table}[tb]
\caption{Equations of the phase boundaries for $J=J_1 > 0$,
$K_1=K_2 \equiv K$, and
$\Delta=\Delta_1$ (corresponding to the top panel of Fig.\
\ref{Fig-2}). The abbreviations $q$ and $\overline{q}$ are
introduced in Eq.~(\ref{eq:qqAbb}). }
\begin{ruledtabular}
\begin{tabular}{c c l}
\hline Phase 1 & Phase 2 & Boundary \\ \hline AF4 & SP  &
$h=\frac{1}{8}\left(4-\Delta+2 \kappa+ q(-\kappa) \right)$
\\
AF4 & F4  & $h=\frac{1}{4}\left(-3 \Delta+ q(-\kappa) \right)$ \\
F4 & SP  & $h=\frac{1}{2}\left(2+\Delta+\kappa \right)$ \\
SM2 & F4  & $h=\frac{1}{4}\left(-3\Delta+4 \kappa+ q(\kappa)\right)$ \\
AF3 & SM2  & $h=\frac{1}{4}\left(-3\Delta-2+ q(-\kappa)\right)$ \\
SM2 & F1  & $h=\frac{1}{4}\left(8+\Delta_1-4\kappa+  q(-\kappa)\right)$ \\
AF3 & F1  & $h=\frac{1}{4}\left(4-\Delta-2 \kappa q(-\kappa)\right)$ \\
F1 & SP  & $h=\frac{3}{2}\kappa$ \\
F4 & F1   & $\kappa=1+\frac{1}{2}\Delta$ \\ \hline
\end{tabular}
\label{Tab:A1}
\end{ruledtabular}
\end{table}

\begin{table}[tb]
\caption{Equations of the phase boundaries for $J=J_1 > 0$,
$K_1=K_2 \equiv K$, and
$\Delta > \Delta_1$ (corresponding to the middle panel of Fig.\
\ref{Fig-2}).}
\begin{ruledtabular}
\begin{tabular}{c c l}
\hline Phase 1 & Phase 2 & Boundary \\ \hline AF4 & SP
&$h=\frac{1}{8}\left(4-\Delta_1+2 \kappa+q(-\kappa) \right)$
\\
AF4 & F3  & $h=\frac{1}{4}\left(q(-\kappa)-q(0) \right)$ \\
F3 & SP  & $h=\frac{1}{4}\left(4-\Delta_1+2 \kappa+q(0) \right)$ \\
AF3 & F3  & $h=\frac{1}{4}\left(2 \kappa-q(0)+ q(\kappa)\right)$ \\
AF3 & F1  & $h=\frac{1}{4}\left(4-\Delta_1-2 \kappa+ q(\kappa)\right)$ \\
AF3 & SM2  & $h=\frac{1}{4}\left(-3\Delta_1+ q(\kappa)\right)$ \\
SM2 & F3  & $h=\frac{1}{4}\left(3\Delta_1+4 \kappa-2q(0)+ q(\kappa)\right)$ \\
SM2 & F1  & $h=\frac{1}{4}\left(8+\Delta_1-4 \kappa+ q(\kappa)\right)$ \\
F1 & SP   & $h=\frac{3}{2} \kappa$ \\
F3 & F1   & $\kappa=1-\frac{1}{4}\left(
\Delta_1-q(0)\right)$\\\hline
\end{tabular}
\label{Tab:A2}
\end{ruledtabular}
\end{table}

\begin{table}[tb]
\caption{Additional equations of the phase boundaries for $J=J_1 >
0$, $K_1=K_2 \equiv K$, and
$\Delta_1 > \Delta$ (corresponding to the lower panel of Fig.\
\ref{Fig-2}). The other equations of phase boundaries are the same
as in Table \ref{Tab:A2}.}
\begin{ruledtabular}
\begin{tabular}{c c l}
\hline Phase 1 & Phase 2 & Boundary \\ \hline
AF4 & F2  & $h=\frac{1}{4}\left(-3\Delta_1+q(-\kappa) \right)$ \\
F2 & SP  & $h=\frac{1}{2}\left(2+\Delta_1+\kappa\right)$ \\
F2 & AF1  & $h=\frac{1}{2}\kappa$ \\
AF1 & SM2  & $h=\frac{1}{4}\left(3\Delta_1-q(\kappa) \right)$ \\
SM2 & F2  & $h=\frac{1}{4}\left(-3\Delta_1+4\kappa+q(\kappa) \right)$ \\
F1 & F2  & $\kappa=1+\frac{1}{4}\Delta_1$ \\
\hline
\end{tabular}
\label{Tab:A3}
\end{ruledtabular}
\end{table}

Analyzing $T=0$ properties of the system under consideration one can find a rich variety of possible ground states.
First of all, one can distinguish cases of ferromagnetic and antiferromagnetic $J$.
Hereafter, we will measure all coupling constants in units of $|J|$.

In the antiferromagnetic region $J > 0$ we observe a strong dependence
on the ratio of
$\Delta/\Delta_1$. For equal anisotropy in the triangles ($\Delta=\Delta_1$),
the system exhibits an exceptional macroscopically degenerate ferrimagnetic ground state with $M=1/2$,
$|\text{F4}\rangle$.
The top panel of Fig.~\ref{Fig-2} presents the phase diagram for
$J > 0$ and $\Delta=\Delta_1$. This phase diagram has a point ($K=0$, $H=0$)
where four ground states merge, with a very high (eight-fold) degeneracy per block.
However, the origin of this degeneracy is trivial, as the system at such values of parameters
is just a set of triangles and single spins decoupled from each other.  One also observes
antiferromagnetic ($M=0$) and ferrimagnetic spatially modulated ($M=1/2$) ground states.
The equations of the corresponding phase boundaries are listed in Table \ref{Tab:A1}.
Here and below we use the following notations:
\begin{eqnarray}
q(x) &=& \sqrt{8\Delta^2+(\Delta_1+2x)^2} \, , \nonumber \\
\overline{q}(x) &=& \sqrt{8\Delta^2+(\Delta+2x)^2} \, , \nonumber \\
q_{\kappa}(x)&=&\sqrt{8 \Delta^2 \kappa^2+(\Delta_1+2 x)^2} \, ,\nonumber \\
\overline{q_{\kappa}}(x)&=&\sqrt{8 \Delta^2 \kappa^2+(\Delta+2 x)^2}
\, .
 \label{eq:qqAbb}
\end{eqnarray}
Throughout this section we use the abbreviations $\kappa=K/|J|$ and $h=H/|J|$.

The middle panel of Fig.\ \ref{Fig-2} presents the phase diagram for
$\Delta/\Delta_1>1$. Depending on the value of the ratio
$\kappa=K/J$ and the magnetic field measured in units of $J$
($h=H/J$), the system exhibits two antiferromagnetic, two
ferrimagnetic, and one spin-modulated ground state, apart from the
spin-polarized one. The present ground-state
phase diagram differs from the previous one (see Fig.\ \ref{Fig-2},
top panel). Here, the degenerate ferrimagnetic ground state
$|\text{F4}\rangle$ is replaced by $|\text{F3}\rangle$ and the
region corresponding to $|\text{SM2}\rangle$ is smaller than in the
previous case. An interesting feature of the symmetry breaking
imposed by the distinct values of the anisotropy is the
disappearance of the points where four phases merge. The equations
for the corresponding phase boundaries can be found in Table
\ref{Tab:A2}.

The lower panel of Fig.\ \ref{Fig-2} presents the phase diagram
for the opposite relation between anisotropy constants $\Delta/\Delta_1<1$
and antiferromagnetic $J>0$. Here one new feature arises.
Namely, one observes the macroscopically
degenerate antiferromagnetic  ground state $|\text{AF1}\rangle$
between the $|\text{F2}\rangle$ and $|\text{SM2}\rangle$ states.
Table \ref{Tab:A3} contains the equations for all phase boundaries.

\begin{figure}[tb!]
\begin{center}
\includegraphics[width=0.9\columnwidth]{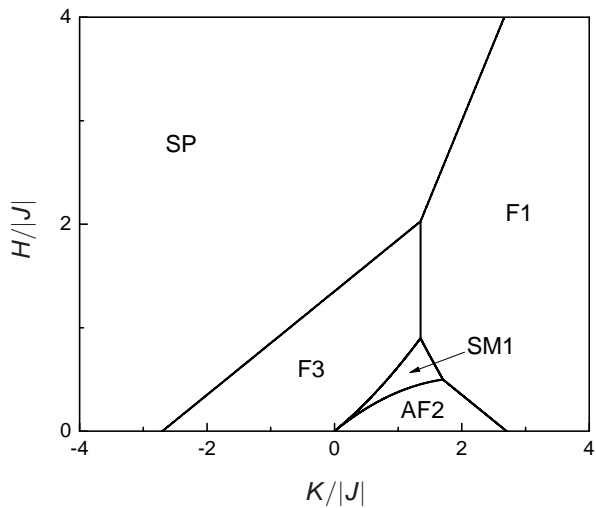}
\end{center}
\caption{\label{Fig-3} Ground-state phase diagram for the case
of ferromagnetic $J_1=J<0$, $K_1=K_2 \equiv K$,  and $\Delta=2$,
$\Delta_1=3$ ($XY$-like anisotropy).
}
\end{figure}

\begin{table}[tb!]
\caption{Equations of phase boundaries for ferromagnetic $J=J_1<0$
and $K_1=K_2 \equiv K$ (Fig.\ \ref{Fig-3}).}
\begin{ruledtabular}
\begin{tabular}{c c l}
\hline Phase 1 & Phase 2 & Boundary \\ \hline
F3 & SP  & $h=\frac{1}{4}\left(\Delta_1-4+2\kappa+q(0) \right)$ \\
AF2 & SM1  & $h=\frac{1}{4}\left(2\kappa+2q(0)-q(-\kappa)-q(\kappa) \right)$ \\
SM1 & F3  & $h=\frac{1}{4}\left(2\kappa-2q(0)+q(-\kappa)+q(\kappa) \right)$ \\
SM1 & F1  & $h=\frac{1}{4}\left(-6\kappa+2\Delta_1-8+q(-\kappa)+q(\kappa) \right)$ \\
AF2 & F1  & $h=\frac{1}{4}\left(\Delta_1-4-2\kappa+q(0) \right)$ \\
F1 & SP  & $h=\frac{3}{2}\kappa$ \\
\hline
\end{tabular}
\label{Tab:A4}
\end{ruledtabular}
\end{table}

Finally, for ferromagnetic $J<0$ one finds the phase diagram shown in
Fig.\ \ref{Fig-3}. The corresponding equations of phase boundaries are
listed in Table \ref{Tab:A4}.

All phase transitions in Figs.\
\ref{Fig-2} and \ref{Fig-3} correspond to level crossings and
thus should be considered as first-order transitions.

\subsection{$J_1=K_1\equiv \tilde{J}$, $J=K_2\equiv K$}

\begin{figure}[tb!]
\begin{center}
\includegraphics[width=0.9\columnwidth]{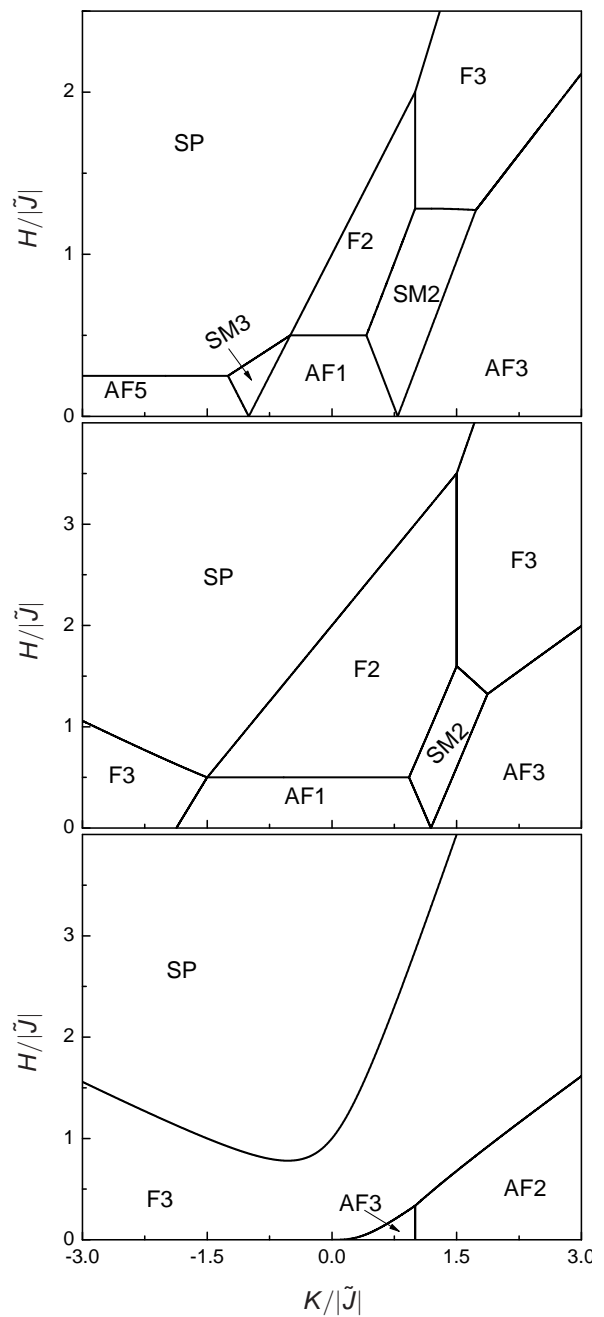}
\end{center}
\caption{\label{Fig-4}Ground-state phase diagrams for the case
$J=K_2\equiv K$, $J_1=K_1\equiv \tilde{J}$, which corresponds most closely
to the fully quantum orthogonal-dimer chain. The top panel
corresponds to isotropic interactions inside the triangles,
$\Delta=\Delta_1=1$, and antiferromagnetic $\tilde{J}>0$.
The middle (lower) panel shows the case of
antiferromagnetic (ferromagnetic) $\tilde{J}$ for the anisotropic case
$\Delta=2$, $\Delta_1=3$ ($XY$-like anisotropy).
}
\end{figure}

The case where dimers have one coupling constant $J_1=K_1\equiv \tilde{J}$ and
diagonal bonds another one $J=K_2\equiv K$ is the closest analog of the
conventional purely quantum orthogonal-dimer chain. The ground state phase
diagrams for this case are presented in Fig.\ \ref{Fig-4}.

\begin{table}[tb!]
\caption{Equations of the phase boundaries for the case
$J_1=K_1\equiv \tilde{J}>0$, $J=K_2\equiv K$ and isotropic interaction inside
triangles $\Delta=\Delta_1=1$ (Fig.~\ref{Fig-4}, top panel).}
\begin{ruledtabular}
\begin{tabular}{c c l}
\hline Phase 1 & Phase 2 & Boundary \\ \hline AF5 & SP
&$h=\frac{1}{4}$
\\
AF5 & SM3  & $h=-\frac{1}{2}\left(1+\Delta+2\kappa\right)$ \\
SM3 & SP  & $h=\frac{1}{6}\left(3+\Delta+2\kappa\right)$ \\
AF1 & SM3  & $h=\frac{1}{2}\left(1+\Delta+2\kappa\right)$ \\
AF1 & F2  & $h=\frac{1}{2}$ \\
F2  & SP  & $h=\frac{1}{2}\left(1+\Delta+2\kappa\right)$ \\
AF1 & SM2  & $h=\frac{1}{4}\left(2+3\Delta-2\kappa-\overline{q_{\kappa}}(\kappa)\right)$ \\
SM2 & F2  & $h=\frac{1}{4}\left(2-3\Delta+2 \kappa+\overline{q_{\kappa}}(\kappa)\right)$ \\
AF3 & SM2   & $h=\frac{1}{4}\left(-2-3\Delta+2\kappa+\overline{q_{\kappa}}(\kappa)\right)$ \\
SM2 & F3
&$h=\frac{1}{4}\left(2+3\Delta+2\kappa-2\Delta\overline{q_{\kappa}}(0)+\overline{q_{\kappa}}(\kappa)\right)$\\
SM2 & F2 & $h=\frac{1}{4}\left(2-3\Delta+2
\kappa+\overline{q_{\kappa}}(\kappa)\right)$\\
F3 & F2 & $\kappa=1$\\
AF3 & F3 &
$h=\frac{1}{4}\left(2\kappa-\Delta\overline{q_{\kappa}}(0)+\overline{q_{\kappa}}(\kappa)\right)$\\
F3 & SP &
$h=\frac{1}{4}\left(2-\Delta+4\kappa+\Delta\overline{q_{\kappa}}(0)\right)$\\
\hline
\end{tabular}
\label{Tab:B1}
\end{ruledtabular}
\end{table}

As before, one should distinguish ferromagnetic and
antiferromagnetic regions of $\tilde{J}$. In both cases, the
appearance of different ground states is affected by the ratio
$\Delta/\Delta_1$. The richest phase diagram exhibiting eight ground
states corresponds to the isotropic case $\Delta=\Delta_1$ and is
shown in the top panel of Fig.\ \ref{Fig-4}. One observes a broad
region corresponding to the macroscopically degenerate
antiferromagnetic ground state $|\text{AF1}\rangle$ which is the
closest analog of the exact dimerized ground state of the purely
quantum orthogonal-dimer chain.\cite{odc,odc2,odc3} The system also
exhibits four ground states with period doubling,
$|\text{AF3}\rangle$, $|\text{AF5}\rangle$, $|\text{SM2}\rangle$,
and $|\text{SM3}\rangle$. It is worth mentioning that there is a
point where the four ground states $|\text{SM3}\rangle$,
$|\text{AF1}\rangle$, $|\text{F2}\rangle$, and $|\text{SP}\rangle$
become degenerate, which implies a large entropy accumulation. For
the isotropic case ($\Delta=\Delta_1=1$) presented in the top panel
of Fig.\ \ref{Fig-4}, the position of this point is $K/\tilde{J}=-1/2$,
$H/\tilde{J}=1/2$ and the corresponding entropy per unit cell is
$S/L=1.44364$. The equations of the phase boundaries can be found in
Table \ref{Tab:B1}.
Throughout this section we use the abbreviations $\kappa=K/|\tilde{J}|$ and $h=H/|\tilde{J}|$.

\begin{table}[tb!]
\caption{Equations of the phase boundaries for the case
$J_1=K_1\equiv \tilde{J}>0$, $J=K_2\equiv K$, and $\Delta=2$,
$\Delta_1=3$ (Fig.~\ref{Fig-4} middle panel).}
\begin{ruledtabular}
\begin{tabular}{c c l}
\hline
Phase 1 & Phase 2 & Boundary \\
\hline
F3 & SP & $h=\frac{1}{4}\left(4\kappa-\Delta_1+2+q_{\kappa}(0) \right)$\\
F3 & AF1  & $h=\frac{1}{4}\left(2+3\Delta_1-q_{\kappa}(0) \right)$ \\
AF1 & F2  & $h=\frac{1}{2}$ \\
F2 & SP  & $h=\frac{1}{4}\left(1+\Delta_1+2\kappa \right)$ \\
AF1 & SM2  & $h=\frac{1}{4}\left(2+3\Delta_1-2\kappa-q_{\kappa}(\kappa) \right)$ \\
SM2 & F2  & $h=\frac{1}{4}\left(2-3\Delta_1+2\kappa+q_{\kappa}(\kappa) \right)$ \\
SM2 & AF3   & $h=\frac{1}{4}\left(-2-3\Delta_1+2\kappa+q_{\kappa}(\kappa) \right)$ \\
SM2 & F3 & $h=\frac{1}{4}\left(2+3\Delta_1+2\kappa-2q_{\kappa}(0)+q_{\kappa}(\kappa) \right)$\\
AF3 & F3 & $h=\frac{1}{4}\left(2\kappa-q_{\kappa}(0)+q_{\kappa}(\kappa) \right)$\\
F3 & SP & $h=\frac{1}{4}\left(1-\Delta_1+5\kappa+q_{\kappa}(0)\right)$\\
F2 & F3 & $\kappa={\Delta_1}/{\Delta}$\\ \hline
\end{tabular}
\label{Tab:B2}
\end{ruledtabular}
\end{table}

The structure of the phase diagram for anisotropic interactions
and antiferromagnetic $\tilde{J}$, $K>0$ is qualitatively similar.
The middle panel of Fig.\ \ref{Fig-4} shows the ground state phase
diagram for $\Delta=2$, $\Delta_1=3$, and $\tilde{J}>0$.
Here one can see
that, in comparison with the isotropic case, there is no region
corresponding to the $|\text{SM3}\rangle$ state and the antiferromagnetic
$|\text{AF5}\rangle$ ground state is replaced by the ferrimagnetic state
$|\text{F3}\rangle$. Still, there is a point of large entropy
accumulation where four ground states $|\text{F3}\rangle$,
$|\text{AF1}\rangle$, $|\text{F2}\rangle$, and $|\text{SP}\rangle$
become degenerate.
In the symmetric but anisotropic case $\Delta=\Delta_1\neq1$,
the value of the entropy at the
corresponding point is $S/L=1.60944$, corresponding to a five-fold
degeneracy per block. Note also the line of degeneracy
between the $|\text{F2}\rangle$ and $|\text{F3}\rangle$ states which one can
see both in the top and middle panel of Fig.\ \ref{Fig-4}. This line
is at $K/\tilde{J}={\Delta_1}/{\Delta}$ and corresponds to the
degenerate $|\text{F4}\rangle$ ground state which in some sense is
a superposition of $|\text{F2}\rangle$ and $|\text{F3}\rangle$.

\begin{table}[tb!]
\caption{Equations of phase boundaries for ferromagnetic
$J_1=K_1\equiv \tilde{J}<0$ and $J=K_2\equiv K$ (lower panel of Fig.\ \ref{Fig-4}).}
\begin{ruledtabular}
\begin{tabular}{c c l}
\hline Phase 1 & Phase 2 & Boundary \\ \hline
F3 & SP  & $h=\frac{1}{4}\left(4\kappa+\Delta_1+q_{\kappa}(0) \right)$ \\
AF3 & F3  & $h=\frac{1}{4}\left(2\kappa-q_{\kappa}(0)+q_{\kappa}(-\kappa) \right)$ \\
AF2 & F3 &
$h=\frac{1}{4}\left(2\kappa-q_{\kappa}(0)+q_{\kappa}(-\kappa-1)
\right)$ \\
AF2 & AF3  & $\kappa=\frac{1}{2}\left(\Delta_1-1 \right)$ \\
\hline
\end{tabular}
\label{Tab:B3}
\end{ruledtabular}
\end{table}

Lastly, the region of ferromagnetic $\tilde{J}<0$ is quite different.
The lower panel of Fig.\ \ref{Fig-4} demonstrates the ground-state phase
diagram for ferromagnetic $\tilde{J}$ and asymmetric anisotropic
couplings $\Delta=2$, $\Delta_1=3$. The equations of the phase boundaries
corresponding to the middle and lower panels of Fig.\ \ref{Fig-4}
are listed in Tables \ref{Tab:B2} and \ref{Tab:B3}, respectively.

Again, all phase transitions in Fig.\ \ref{Fig-4}
are first-order transitions in the sense that they
correspond to level crossings.

\section{Magnetization process}

\label{sec5}

The phase diagrams show that the model (\ref{H}), (\ref{Hi})
exhibits a large diversity of magnetic behaviors.
Namely, depending on the values of parameters, the following
step-like transitions between plateaux in the
magnetization curves can be observed when the magnetic field is varied
from zero to the saturation value:
$M=0\rightarrow 1$,
$M=0\rightarrow 1/2\rightarrow 1$,
$M=0\rightarrow 1/4\rightarrow 1/2\rightarrow 1$,
$M=0\rightarrow 1/4\rightarrow 1$,
$M=1/2\rightarrow 1$,
and $M=1/4\rightarrow 1/2\rightarrow 1$.
As there are no excitation bands in the model, all transitions
between different ground states are strictly step-like at $T=0$.

Here we illustrate some variants of the magnetization process and demonstrate
the effect of a finite temperature using the exact solution of the model.
For sufficiently low temperatures, the form of the magnetization
curve almost coincides with the zero-temperature limit.
Further details for $T=0$ will be given in the comparison with the
full quantum mechanical Heisenberg model in Section \ref{sec:HeisMag} below.

\begin{figure}[tb!]
\begin{center}
\includegraphics[width=0.9\columnwidth]{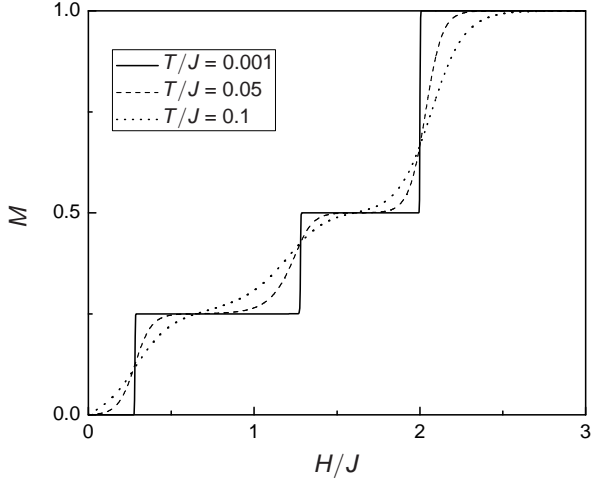}
\end{center}
\caption{\label{Fig-5} Magnetization curve for $J_1=K_1=1$,
$J=K_2=1$, and $\Delta=\Delta_1=1$. The corresponding ground states
can be seen both in the top panel of Fig.\ \ref{Fig-2} and the top
panel of Fig.\ \ref{Fig-4}. For very low temperature (solid line)
the sequence of step-like transitions between the corresponding
ground states is recovered. Dashed and dotted lines correspond to
$T/J=0.05$ and $T/J=0.1$, respectively. }
\end{figure}

\begin{figure}[tb!]
\begin{center}
\includegraphics[width=0.9\columnwidth]{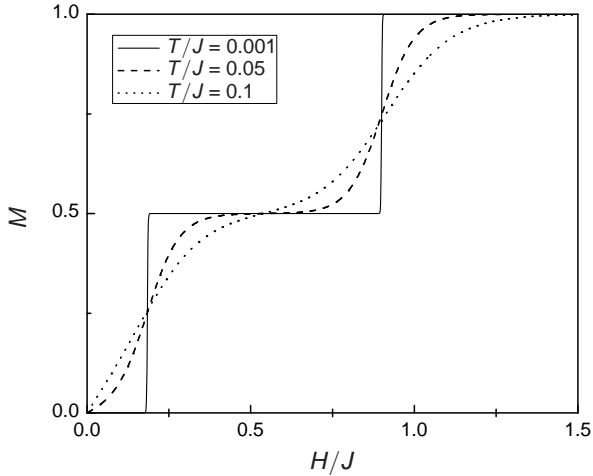}
\end{center}
\caption{\label{Fig-6} Magnetization curve for $J=J_1=1$,
$K_1=K_2=-1$, and $\Delta=0.5$, $\Delta_1=0.8$ (Ising-like
anisotropy). The solid line corresponding to $T/J=0.001$ shows sharp
transitions between three plateaux: $M=0$, $1/2$, and $1$. Dashed
and dotted lines correspond to $T/J=0.05$ and $T/J=0.1$,
respectively. }
\end{figure}

Fig.\ \ref{Fig-5} shows the magnetization curve
for the most symmetric case $J=J_1=K_1=K_2=1$ and $\Delta=\Delta_1=1$,
for $T/J=0.001$, $ 0.05$, and $0.1$. One can see here a sequence of
magnetization plateaux corresponding to the following transitions
between the ground states at $T=0$:
$|\text{AF3}\rangle\rightarrow|\text{SM2}\rangle\rightarrow|\text{F4}\rangle\rightarrow|\text{SP}\rangle$
which corresponds to both phase diagrams from Fig.\ \ref{Fig-2}
and Fig.\ \ref{Fig-4}. The line $K/\tilde{J}=1$ from the upper panel of Fig.\
\ref{Fig-4} where $|\text{F2}\rangle$ and $|\text{F3}\rangle$ become
degenerate corresponds to the degenerate $|\text{F4}\rangle$ ground state.

\begin{figure}[tb!]
\begin{center}
\includegraphics[width=0.9\columnwidth]{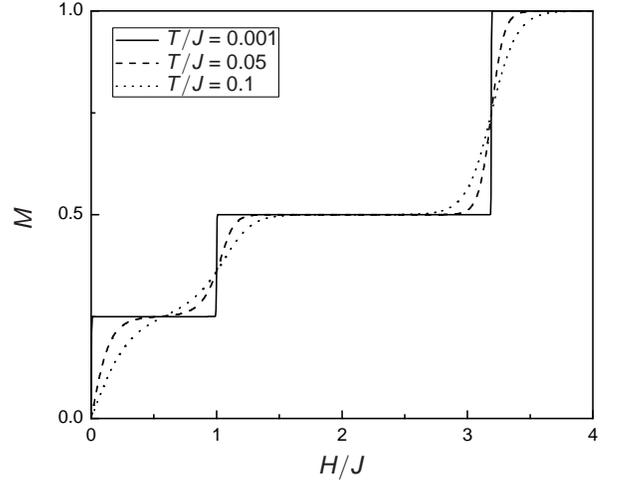}
\end{center}
\caption{\label{Fig-7} Magnetization curve for $J_1=K_1=1$,
$J=K_2=1.19034$, and $\Delta=2$, $\Delta_1=3$ ($XY$-like
anisotropy). The solid line corresponds to $T/J=0.001$ and shows
sharp transitions between three plateaux: $M=1/4$, $1/2$, and $1$.
Dashed and dotted lines correspond to $T/J=0.05$ and $T/J=0.1$,
respectively. }
\end{figure}

Fig.\ \ref{Fig-6} exhibits another
form of magnetic behavior  with
only two intermediate plateaux at $M=0$ and $1/2$, respectively. For
the values of parameters corresponding to the figure ($J=J_1=1$,
$K_1=K_2=-1$ and $\Delta=0.5$, $\Delta_1=0.8$), these plateau correspond
to the $|\text{AF4}\rangle$ and $|\text{F2}\rangle$ ground states (see
lower panel of Fig.\ \ref{Fig-2}).

Fig.\ \ref{Fig-7} shows
the magnetization curves for $J_1=K_1=1$, $J=K_2=1.19034$ and
$\Delta=2$, $\Delta_1=3$ which corresponds to the middle panel of Fig.\
\ref{Fig-4}. This exceptional value of $K/\tilde{J}$ corresponds to the
point of the phase diagram where the three ground states
$|\text{AF1}\rangle$, $|\text{SM2}\rangle$, and $|\text{F3}\rangle$
become degenerate, $(K/\tilde{J}=\frac{1}{16}(-14 + 2 \sqrt{273})$,
$H=0$) and the first plateau in the magnetization  curve corresponds
to $M=1/4$.

\section{Heisenberg model}

\label{sec6}

In the following we would like to test to which extent the reduction to
Ising interactions for certain bonds affects the properties of the model.
For this purpose we will use the Heisenberg-Ising variant of the model
(\ref{H}), (\ref{Hi}), {\it i.e.}, we will put $\Delta = \Delta_1 = 1$.

The comparison will be carried out with the Heisenberg analog of the
Hamiltonian
\begin{eqnarray}
{\cal H} &=& \sum_{i=1}^{N/4} \biggl(
J \, \bmth{S}_{i,1} \cdot \left(\bmth{S}_{i,2} + \bmth{S}_{i,3} \right)
+ J_1 \, \bmth{S}_{i,2} \cdot \bmth{S}_{i,3}
\nonumber \\
&& \qquad
+ K_1 \, \bmth{S}_{i,1} \cdot \bmth{S}_{i,4}
+ K_2 \, \left(\bmth{S}_{i,2} + \bmth{S}_{i,3} \right) \cdot \bmth{S}_{i+1,4}
\nonumber \\
&& \qquad
- H \sum_{a=1}^4 S_{i,a}^z
\biggr) \, .
\label{eq:Hheis}
\end{eqnarray}
Note that we have replaced the Ising spin $\sigma_i$ in Eqs.\ (\ref{H}), (\ref{Hi})
with a full $s=1/2$ Heisenberg spin $\bmth{S}_{i,4}$.

\subsection{Method}

Firstly, one can use conservation of the total $S^z = \sum_{i,a} S_{i,a}^z$
to relate the energy $E(S^z,H)$ in a magnetic field to that at zero
field
\begin{equation}
E(S^z,H) = E(S^z,0) - H\,S^z \, .
\label{eq:ESz}
\end{equation}
It is therefore sufficient to diagonalize (\ref{eq:Hheis}) in
a given sector of $S^z$ for $H=0$.

Secondly, we can introduce composite operators
\begin{equation}
\bmth{T}_i = \bmth{S}_{i,2} + \bmth{S}_{i,3}
\label{eq:DefTi}
\end{equation}
and rewrite the $H=0$ case of (\ref{eq:Hheis}) as follows:
\begin{eqnarray}
{\cal H}_{H=0} &=& \sum_{i=1}^{N/4} \biggl(
J \, \bmth{S}_{i,1} \cdot \bmth{T}_i
+ J_1 \, \left(\bmth{T}_i^2 - \frac{3}{2}\right)
\nonumber \\
&& \qquad
+ K_1 \, \bmth{S}_{i,1} \cdot \bmth{S}_{i,4}
+ K_2 \, \bmth{T}_i \cdot \bmth{S}_{i+1,4}
\biggr) \, .
\label{eq:HheisCons}
\end{eqnarray}
This puts our model into the class of one-dimensional models with local
conservation laws.%
\cite{str,strMCE,roj2,odc,odc2,odc3,HSR04,Gelfand91,TKS96,IvRi97,MTM99,HMT00,%
GVAHW00,TrSe00,HoBr01,ChBo02,SchRi02b,RoAl03,DR04,Richter05,%
DeRi06,DRHS07,ivanov09,DTK10,hida1,hida2,MPSM10,HHPR11,MDR11,HT11}
This property can be used to simplify the diagonalization:
Since the total spin $\bmth{T}_i$ on each vertical dimer is separately
conserved, diagonalization of (\ref{eq:Hheis}) is reduced to
diagonalization of (\ref{eq:HheisCons}) for all possible combinations
$T_i = 0$, $1$ (with $\bmth{T}_i^2 = T_i \, (T_i + 1)$).
In fact, a singlet $T_i=0$ splits the chain into ``fragments'':
all states for a given singlet pattern are product
states of the corresponding eigenstates of the open fragments
separated by these singlets.
Hence, it is sufficient to diagonalize all open fragments
with up to $N/4-1$ blocks in which $T_i = 1$ and one periodic
system with $N/4$ with all $T_i=1$ in order to obtain the
spectrum for a periodic chain with $N$ sites.\cite{DRHS07,HHPR11}
For $K_2 = J$, the use of reflection symmetry allows us
to fully diagonalize a system with $N=24$ spins 1/2, a
task which in view of the large unit cell of the model would
be very demanding if not impossible for the original Hamiltonian
(\ref{eq:Hheis}).

Ground states consist of a periodic repetition of one fragment
of a certain length or a periodic system of a given length
with all $T_i = 1$.\cite{odc3}
In the former case, we can construct the exact thermodynamic limit
of the magnetization curves, provided that we have access to
sufficiently long fragments. For the periodic system, we use
(i) values for plateau boundaries obtained at sufficiently large $N$
and
(ii) the standard midpoint method\cite{BoFi64,BoPa85}
as applied to the largest available system size to approximate
the thermodynamic limit in smooth regions of the magnetization
curve.

The main limitation for the computation of ground-state properties
is the diagonalization of a periodic system with all $T_i=1$.
To push this a bit further, we have used the ALPS~1.3
implementation\cite{alps2} of the density-matrix renormalization
group (DMRG) method.\cite{DMRGa,DMRGb} Below we show results
based on DMRG for $N=40$ in the regime $M<1/2$ and $N=64$
with $m=400$ kept states as well as $N=96$ with $m=600$ kept
states. In the case $J_1=J=1$, $K_1=K_2=-1$, we have exceptionally
increased the number of kept states to $m=600$ and $800$
for $N=64$ and $96$, respectively. We have always also
performed computations with lower $m$ to ensure convergence
of the data.

\subsection{Magnetization curves}

\label{sec:HeisMag}

\begin{figure}[tb!]
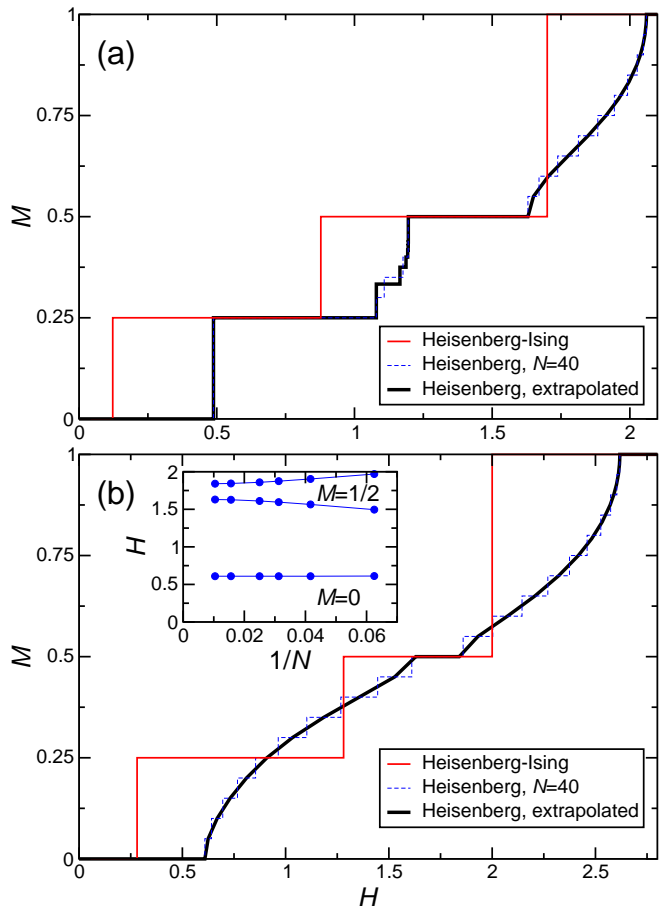

\begin{center}
\includegraphics[width=\columnwidth]{mag_J1=1_J2=0p7_J3=1}
\includegraphics[width=\columnwidth]{mag_J1=1_J2=1_J3=1}
\end{center}
\caption{(Color online) Zero-temperature magnetization curves for
two cases where all exchange constants are antiferromagnetic:
$J_1=K_1=1$, $J=K_2=0.7$ (a) and
$J_1=J=K_1=K_2=1$ (b).
Dashed lines show numerical results for
a Heisenberg chain with $N=40$ spins 1/2 and thick full lines an
extrapolation to the thermodynamic limit. Thin full lines show the
exact solution of the Heisenberg-Ising model in the thermodynamic
limit $N = \infty$.
The inset of panel (b) shows the finite-size behavior of
the edge of the $M=0$ plateau (spin gap) and the two edges
of the $M=1/2$ plateau at $J_1=J=K_1=K_2=1$.
\label{fig:magSet1} }
\end{figure}

Now we present examples of numerical results for magnetization curves
of the Heisenberg model (\ref{eq:Hheis}) and compare them with
the Heisenberg-Ising model (\ref{H}), (\ref{Hi}) with $\Delta = \Delta_1 = 1$.

\subsubsection{$J_1=K_1=1$, $J=K_2=0.7$}

Fig.~\ref{fig:magSet1}(a) shows the case $J_1=K_1=1$, $J=K_2=0.7$
which has been discussed previously.\cite{odc2,odc3,HSR04}
Our new exact diagonalization results agree with the previous ones.
Note that the spin-1/2 Heisenberg model most likely gives rise to an
infinite sequence of magnetization plateaux which accumulate just
below $M=1/2$.\cite{odc3} This infinite sequence of plateaux is not
reproduced by the Heisenberg-Ising model (red thin full line in
Fig.~\ref{fig:magSet1}(a)) which does, however, reproduce some of
the most prominent features: (i) a spin gap at $M=0$, (ii) two broad
plateaux at $M=1/4$ and $1/2$, and (iii) a jump from $M=0$ to $1/4$.
This choice of parameters corresponds to a cut through the
top panel of Fig.~\ref{Fig-4}
at $K/\tilde{J}=0.7$ such that one can read off the corresponding ground
states of the Heisenberg-Ising model.

For $M=0$, the ground state of the Heisenberg model is characterized
by all $T_i =0$. This leaves isolated horizontal dimers which also
form singlets. The corresponding ground state of the Heisenberg-Ising
model is the $|\text{AF1}\rangle$ state. This is in fact the closest
analog of the fully dimerized ground state: the vertical dimers
are also in the singlet state for the $|\text{AF1}\rangle$ state.
However, due to the missing quantum fluctuations the horizontal dimer
bond is unable to form a singlet. The two local classical
antiparallel spin configurations appear both with equal
probability in the $|\text{AF1}\rangle$ state instead of
forming a singlet superposition. This also gives rise to
a macroscopic ground-state degeneracy of the Heisenberg-Ising
model with entropy $S/L=\ln2$ per unit cell.

\begin{table}[tb!]
\caption{Local spin polarizations on the $M=1/4$ plateau
for $J_1=K_1=1$, $J=K_2=0.7$ (see Fig.~\ref{fig:magSet1}(a)).
}
\begin{ruledtabular}
\begin{tabular}{c c c}
\hline
& Heisenberg & Heisenberg-Ising \\ \hline
$S_{i,1}^z$                                & $0.3312$  & $0.5$     \\ 
$S_{i,2}^z + S_{i,3}^z$ ($=T_i^z$)         & $0$       & $0$       \\
$S_{i+1,4}^z$ / $\sigma_{i+1}$             & $0.3312$  & $-0.5$    \\ 
$S_{i+1,1}^z$                              & $-0.2237$ & $-0.1022$ \\ 
$S_{i+1,2}^z + S_{i+1,3}^z$ ($=T_{i+1}^z$) & $0.7849$  & $0.6022$  \\ 
$S_{i+2,4}^z$ / $\sigma_{i+2}$             & $-0.2237$ & $0.5$     \\ 
 \hline
\end{tabular}
\label{Tab:HeisIsingM1o4}
\end{ruledtabular}
\end{table}

For $M=1/4$, $T_i=0$ and $1$ alternate in the ground state of the Heisenberg model,
{\it i.e.}, the ground state breaks translational symmetry spontaneously.
The same translational symmetry breaking is observed in the corresponding
$|\text{SM2}\rangle$ state of the Heisenberg-Ising model. A more
detailed comparison of the two states is given by the local spin
polarizations along the field direction shown in Table~\ref{Tab:HeisIsingM1o4}.
For the Heisenberg model, this follows from the local expectation
values in the $S^z=1$ sector of a fragment consisting of a single
block. For the Heisenberg-Ising model one needs instead the
expectation values in the $|v_{\frac{1}{2},s}^-\rangle$ state
of a single quantum triangle. Using the explicit form of the
eigenvector given in  Eq.\ (\ref{evec}) one arrives at
\begin{eqnarray}
\langle
v_{\frac{1}{2},s}^-|(S_2^z+S_3^z)|v_{\frac{1}{2},s}^-\rangle&=&\frac{c_-^2}{2+c_-^2}\,,
\\
\langle
v_{\frac{1}{2},s}^-|S_1^z|v_{\frac{1}{2},s}^-\rangle&=&\frac{1-\frac{1}{2}c_-^2}{2+c_-^2}\,.
\nonumber \label{eq:exp_val}
\end{eqnarray}
Combination with the $|\text{SM2}\rangle$
wave function given in Eq.~(\ref{eq:SM}) yields the corresponding
entries in Table~\ref{Tab:HeisIsingM1o4}. The entries in
Table~\ref{Tab:HeisIsingM1o4} exhibit a similar structure
of the $M=1/4$ state at $J_1=K_1=1$, $J=K_2=0.7$ in the Heisenberg
and Heisenberg-Ising models. In particular, vertical dimers in
the singlet state alternate with predominantly polarized ones.
The quantitative differences observed in Table~\ref{Tab:HeisIsingM1o4}
can be attributed to the interaction of the Heisenberg-Ising
model breaking the spatial reflection symmetry of the
Heisenberg model.

In the $M=1/2$ state of the Heisenberg model
one finds all vertical dimers in the state $T_i=1$.
This state is non-degenerate and thus translationally invariant.
The spin polarizations in this state have to be determined
numerically; using DMRG one finds
$S_{i,1}^z = S_{i,4}^z = 0.0640$ 
and
$S_{i,2}^z + S_{i,3}^z = T_i^z = 0.8721$. 
\cite{footnote2}
The corresponding ground state of the Heisenberg-Ising
model is the $|\text{F2}\rangle$ state. Accordingly, the
spin polarizations are read off from Eq.~(\ref{eq:F})
as $S_{i,1}^z = \sigma_i = 1/2$ and
$S_{i,2}^z + S_{i,3}^z = 0$. So, in this case the difference
between the Heisenberg model and the Heisenberg-Ising model is
bigger although the $M=1/2$ ground states still have the
same translational symmetry.

\begin{figure}[tb!]
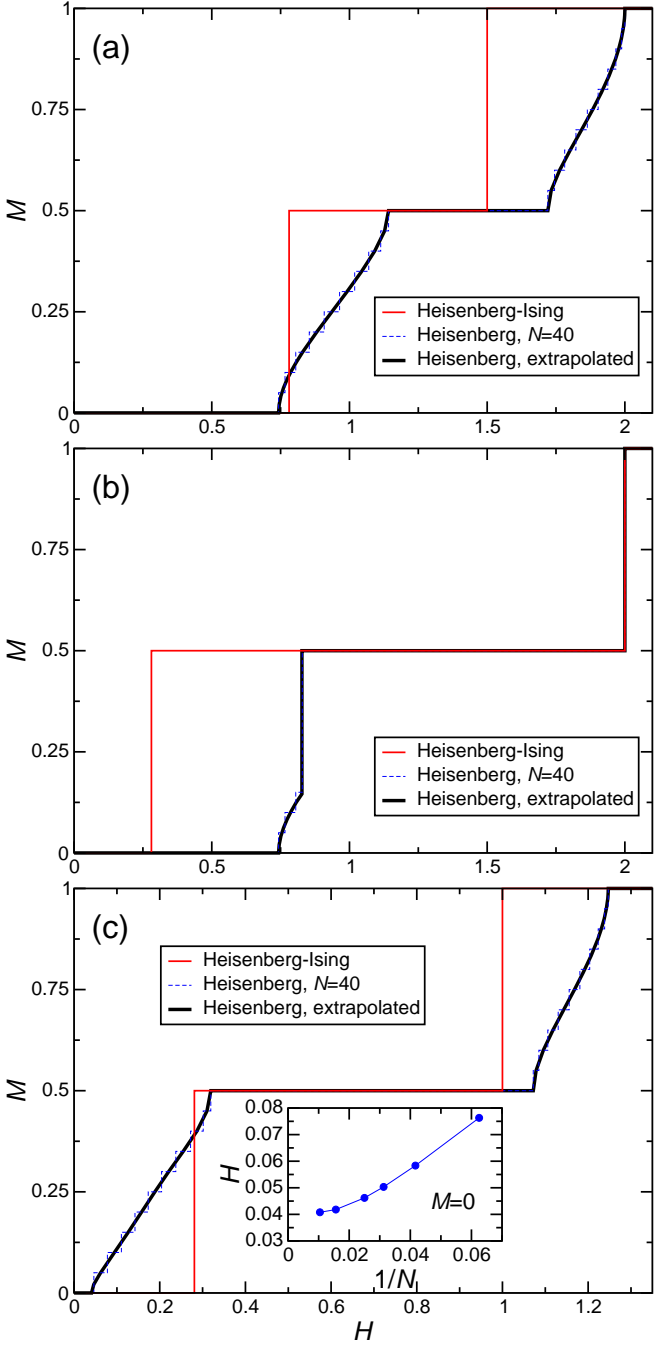

\begin{center}
\includegraphics[width=\columnwidth]{mag_J1=-1_J2=1_J3=-1}
\includegraphics[width=\columnwidth]{mag_J1=1_J2=1_J3=-1}
\includegraphics[width=\columnwidth]{mag_J1=1_J2=1_J3=-1_J4=-1}
\end{center}
\caption{(Color online)
Same as Fig.~\ref{fig:magSet1}, but for cases which
contain ferromagnetic exchange constants, namely
$J_1=K_1=-1$, $J=K_2=1$ (a),
$J_1=J=K_2=1$, $K_1=-1$ (b), and
$J_1=J=1$, $K_1=K_2=-1$ (c).
The inset of panel (c) shows the finite-size behavior of
the spin gap (edge of the $M=0$ plateau)
at $J_1=J=1$, $K_1=K_2=-1$.
\label{fig:magSet2} }
\end{figure}

\begin{table*}[tb!]
\caption{Values of local spin polarizations for the ground states
corresponding to the magnetization plateaux
in Fig.\ \ref{fig:magSet2}. The $|\text{AF4}\rangle$
ground state breaks translational symmetry such that blocks with
different signs alternate. Hence, the upper (lower) sign should be
taken for an even (odd) cell of the $|\text{AF4}\rangle$ state.}
\begin{ruledtabular}
\begin{tabular}{l|cccc|ccc}
\hline
Case & \multicolumn{4}{c|}{Heisenberg-Ising} & \multicolumn{3}{c}{Heisenberg} \\
& Ground-state label & $S_{i,1}^z$ & $\sigma_i$ & $S_{i,2}^z+S_{i,3}^z$
& $S_{i,1}^z$ & $S_{i,4}^z$ & $T_i^z = S_{i,2}^z+S_{i,3}^z$
\\ \hline
Figs.\ \ref{fig:magSet2}(a),(b), $M=0$
  & AF2 & $-0.4621$  & $1/2$ & $0.9621$
        & $-$ & $-$ & $-$ \\
Fig.\ \ref{fig:magSet2}(c), $M=0$
  & AF4 & $\mp1/6$    & $\pm1/2$ & $\pm2/3$
        & $-$ & $-$ & $-$ \\
Fig.\ \ref{fig:magSet2}(a), $M=1/2$
  & F3  & $-1/6$    & $1/2$    & $2/3$
       & $0.2640$ & $0.2640$ & $0.4720$ \cite{footnote2} \\
Fig.\ \ref{fig:magSet2}(b), $M=1/2$
  & F2  & $1/2$ & $1/2$ & $0$
        & $1/2$ & $1/2$ & $0$ \\
Fig.\ \ref{fig:magSet2}(c), $M=1/2$
  & F4  & $1/6$    & $1/2$    & $1/3$
         & $-0.0799$ & $0.4287$ & $0.6512$ \cite{footnote2} \\
 \hline
\end{tabular}
\label{Tab:C}
\end{ruledtabular}
\end{table*}

\subsubsection{$J_1=J=K_1=K_2=1$}

The effect of increasing $J=K_2$ from $0.7$ to $1$ is
demonstrated in Fig.~\ref{fig:magSet1}(b). The magnetization curve of the
Heisenberg-Ising-model (which was in fact already presented in
Fig.\ \ref{Fig-5}) remains similar to the one in Fig.~\ref{fig:magSet1}(a).
However, some of the plateau states of the
Heisenberg-Ising-model change; the sequence is now
$|\text{AF3}\rangle\rightarrow|\text{SM2}\rangle\rightarrow|\text{F3}\rangle\rightarrow|\text{SP}\rangle$
(see top panel of Fig.~\ref{Fig-4}).
The Heisenberg model still has a spin gap at
$M=0$, but its state is also changed:
now it arises out of a periodic system with all $T_i=1$,
as already observed in Ref.~\onlinecite{SchRi02b}. The $M=1/2$ state
of the Heisenberg model also corresponds to all $T_i=1$ and
is non-degenerate.
Since in this case finite-size
effects are important at $M=1/2$, we have performed a finite-size
analysis which is shown in the inset of Fig.~\ref{fig:magSet1}(b).
While the width of the $M=1/2$ plateau initially shrinks with growing
$N$, $N=96$ is a good approximation to the thermodynamic limit
$1/N=0$. In particular, we conclude that the $M=1/2$ plateau
survives for $N=\infty$. All further plateaux in the
magnetization curve of the Heisenberg model are probably gone.

\subsubsection{$K_1 < 0$}

Fig.~\ref{fig:magSet2} shows three examples of magnetization curves
with ferromagnetic $K_1 < 0$ and potentially a further ferromagnetic
exchange constant. In all three cases shown in
Fig.~\ref{fig:magSet2}, the Heisenberg-Ising model has plateaux at
$M=0$ and $1/2$, separated by jumps in the magnetization curve.
These two plateaux are also present in the Heisenberg model.
However, in the case $J_1=J=1$ and $K_1=K_2=-1$ a finite-size
analysis is again necessary to show that the spin
gap (corresponding to the $M=0$ plateau) persists in the
thermodynamic limit. According to the data shown in the
inset of Fig.~\ref{fig:magSet2}(c), we expect $N=96$ to be
a good approximation to the thermodynamic limit even if
the spin gap decreases substantially with growing $N$ for
the smallest values of $N$ considered.

At a quantitative level, the spin gap of the Heisenberg model compares
favorably to that of the Heisenberg-Ising model in the case of
Fig.~\ref{fig:magSet2}(a) and the width of the $M=1/2$ plateau compares
favorably in all three cases presented in Fig.~\ref{fig:magSet2}. The best
agreement is observed for the case presented in Fig.~\ref{fig:magSet2}(b)
($J_1=J=1$, $K_1=K_2=-1$). In this case, the jump between $M=1/2$ and $1$
is reproduced exactly and both models give rise to a jump at the lower
edge of the $M=1/2$ plateau, albeit not exactly at the same magnetic
field.

\begin{figure}[tb!]
\begin{center}
\includegraphics[width=\columnwidth]{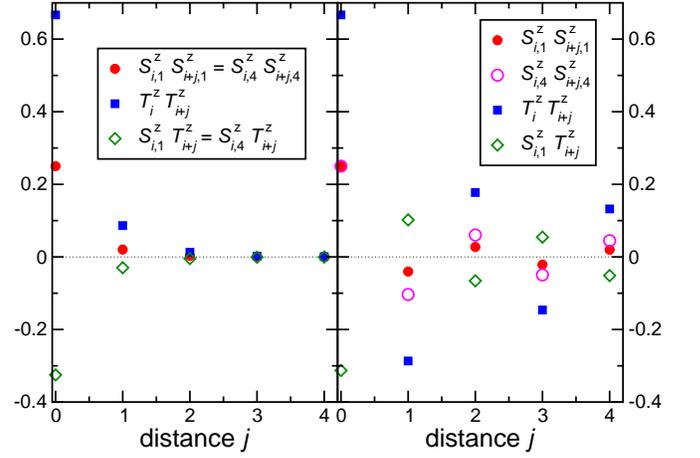}
\end{center}
\caption{(Color online)
Correlation functions of the $z$-components for the Heisenberg
model with $N=32$ sites at $M=0$. $j$ measures the distance in units of
cells; due to the periodic boundary conditions, the maximum
accessible distance is $N/8 = 4$. The left panel is for
$J_1 = \pm 1$, $J=K_2=1$, $K_1=-1$ whereas the right panel shows results for
$J_1=J=1$, $K_1=K_2=-1$.
\label{fig:cfns} }
\end{figure}

For the three examples shown in Fig.~\ref{fig:magSet2}, the spin gap
of the Heisenberg model always arises out of a periodic pattern
$T_i=1$. This conclusion agrees with Ref.~\onlinecite{SchRi02b} in
the case of Figs.~\ref{fig:magSet2}(a) and (b). In fact, the behavior
at low magnetic fields is identical in these two cases, including
the feature at low magnetizations in Fig.~\ref{fig:magSet2}(b).
To the best of our
knowledge, the case presented in Fig.~\ref{fig:magSet2}(c) has not
been discussed previously. The ground state of the Heisenberg-Ising
system at $M=0$ is the $|\text{AF2}\rangle$ one in the case of
Figs.~\ref{fig:magSet2}(a) and (b) and the $|\text{AF4}\rangle$ one
in the case of Fig.~\ref{fig:magSet2}(c). The corresponding local
spin expectation values are given in Table \ref{Tab:C}. In the
Heisenberg model at zero magnetization, individual spin expectation
values vanish for symmetry reasons. Hence, for the Heisenberg model
at $M=0$, we have to look at correlation functions in order to
determine the structure of the ground state. Fig.\ \ref{fig:cfns}
shows selected correlation functions computed by exact
diagonalization for a periodic $N=32$ system as a function of
distance $j$ between cells. The correlations of the left panel
reveal a structure which is similar to the $|\text{AF2}\rangle$
state, {\it i.e.}, the spins $S_{i,1}^z$ and $S_{i,4}^z$ next to the
composite spin $T_i^z$ point in the opposite direction of this
composite spin while spins of the same type show a tendency of
aligning parallel between different cells. By contrast, in the right
panel of  Fig.\ \ref{fig:cfns} one observes a tendency of {\em
antiparallel} alignment between cells at an odd distance while the
internal structure of a cell is the same as before, as expected for
the $|\text{AF4}\rangle$ state. In the case of the left panel of
Fig.\ \ref{fig:cfns}, correlations are short-ranged while for the
right panel they decay more slowly in accordance with the size of
the spin gap (given by the width of the $M=0$ plateau) in
Figs.~\ref{fig:magSet2}(a),(b) and Fig.~\ref{fig:magSet2}(c),
respectively. Overall, we conclude that for all three cases
considered in Fig.~\ref{fig:magSet2}, the $M=0$ ground state of the
Heisenberg-Ising model is a good qualitative representation of the
$M=0$ ground state of the Heisenberg model.

Turning to finite magnetizations, we observe that
in the case of  Figs.~\ref{fig:magSet2}(a) and (c), the $M=1/2$ plateau state
of the Heisenberg model
arises out of a system with all $T_i =1$ while in the case of
Fig.~\ref{fig:magSet2}(b) it is characterized by all $T_i=0$.
The corresponding
local spin expectation values are given in Table \ref{Tab:C}.
The ground states of the Heisenberg-Ising model corresponding to
the $M=1/2$ plateau in the three cases
presented in  Figs.~\ref{fig:magSet2}(a), (b), and (c) are
$|\text{F3}\rangle$, $|\text{F2}\rangle$, and $|\text{F4}\rangle$, respectively.
The spin expectation values for these
states are included in Table \ref{Tab:C}. Note that the $|\text{F4}\rangle$
state is macroscopically degenerate (see Eq.~(\ref{eq:F4})) such that
the average over this manifold yields effectively the average of the local
expectation values in the $|\text{F2}\rangle$ and $|\text{F3}\rangle$
states. In the case of Fig.~\ref{fig:magSet2}(a), the reflection-symmetry
breaking interactions of the Heisenberg-Ising model lead to a breaking
of reflection symmetry in the $M=1/2$ state which is not present in the
Heisenberg model. Otherwise, the structure of the $M=1/2$ states of the
Heisenberg-Ising and Heisenberg model is at least similar in the
case of Figs.~\ref{fig:magSet2}(a) and (c). In the case
of Fig.~\ref{fig:magSet2}(b) we even find an identical $M=1/2$ state
as reflected by identical local spin expectation
values for the Heisenberg-Ising and Heisenberg model.

\subsection{Entropy and cooling rate}

We now turn to thermodynamic properties. Since one of the most important
properties of the present type of models is the magnetocaloric effect,%
\cite{strMCE,ZhHo04,DR04,ZhiTsu04,ZhiTsu05,Richter05,DeRi06,SSHSR06,DRHS07,RDH08,HHPR11}
we focus on the entropy $S$ as a function of magnetic field $H$ and
temperature $T$ as well as the adiabatic cooling rate as a function
of magnetic field. These quantities have been computed for the
Heisenberg model by full diagonalization.

The orthogonal arrangement of dimers in our model is very similar to the
two-dimensional Shastry-Sutherland model.\cite{ShaSu81} The latter model
is found to have a good realization in SrCu$_2$(BO$_3$)$_2$ with a ratio
between dimer and plaquette couplings estimated to be close to $0.7$ (see
Ref.~\onlinecite{MiUe03} for a review). Hence, the case $J_1=K_1=1$,
$J=K_2=0.7$ can be viewed as the one-dimensional counterpart of the
two-dimensional Shastry-Sutherland model for SrCu$_2$(BO$_3$)$_2$. Indeed,
in this parameter regime, both the one- and two-dimensional Heisenberg
model have an exact dimer ground state (compare section \ref{sec:HeisMag}
and Ref.~\onlinecite{ShaSu81}, respectively) and it was noted early on
that the magnetization curve of SrCu$_2$(BO$_3$)$_2$ increases very steeply
above a magnetic field $H \approx 20$T, like a system of isolated
dimers\cite{KZSMOKKSGU99} or the one-dimensional model, see
Fig.~\ref{fig:magSet1}(a).

\begin{figure}[tb!]
\begin{center}
\includegraphics[width=\columnwidth]{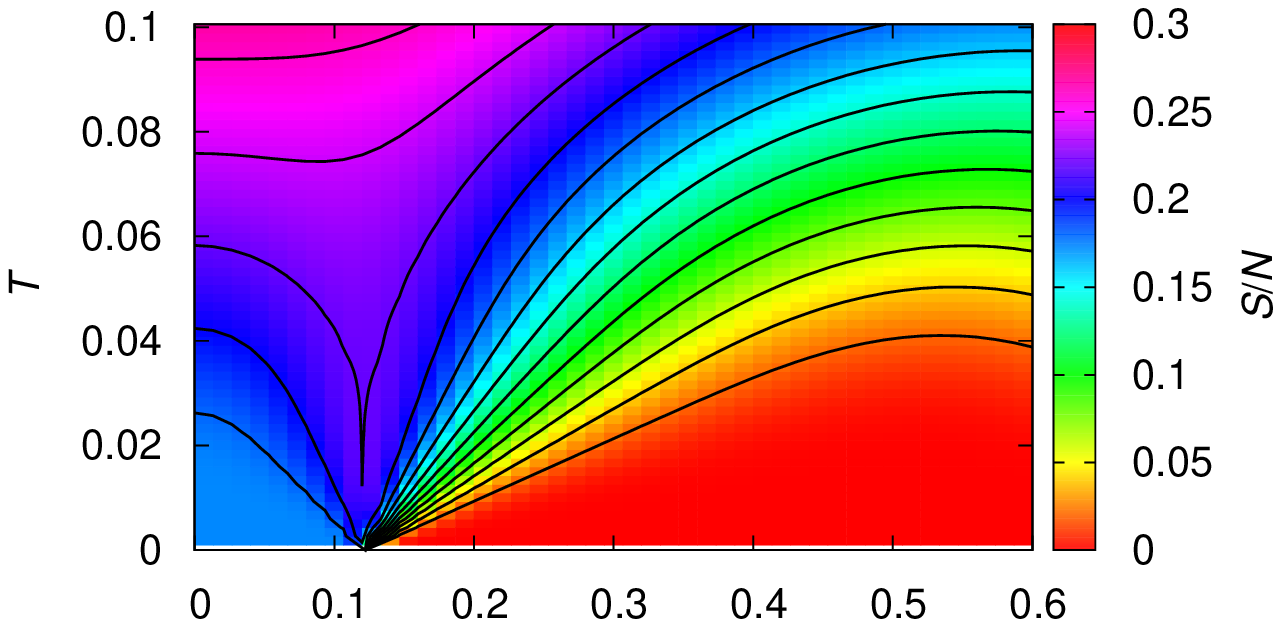}
\includegraphics[width=\columnwidth]{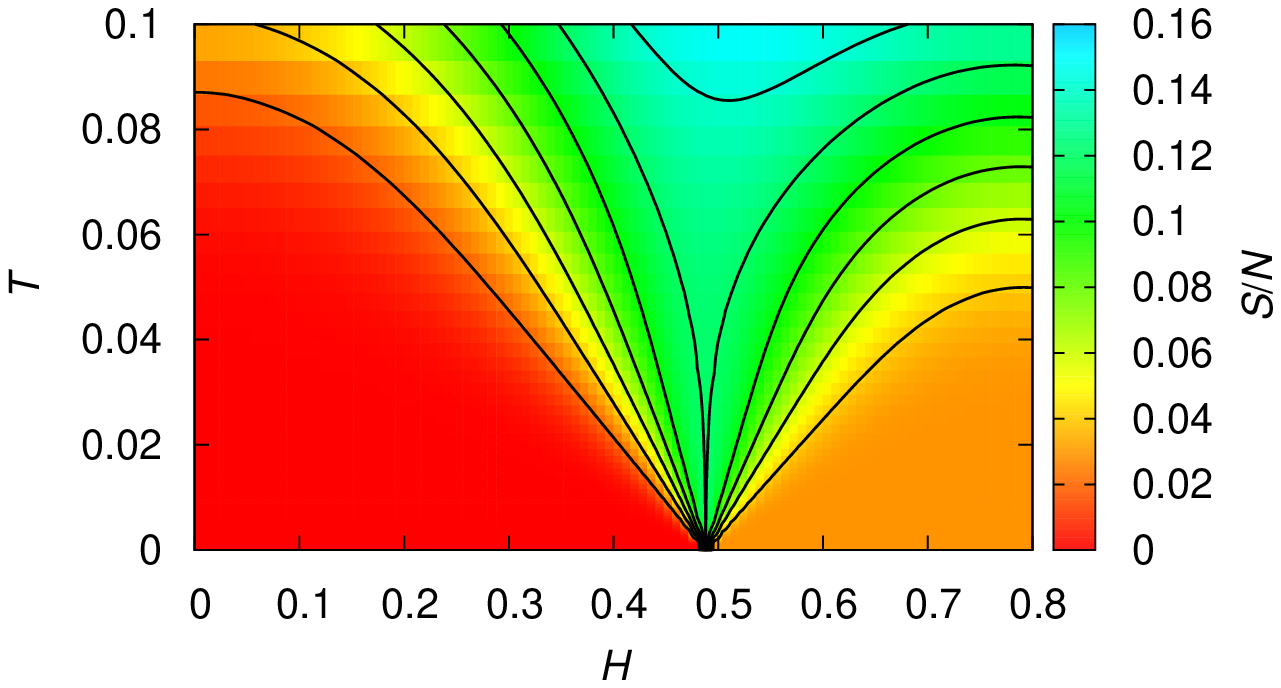}
\end{center}
\caption{(Color online) Entropy per spin $S/N$
as a function of $H$ and $T$ for $J_1=K_1=1$, $J=K_2=0.7$.
The top panel shows results for the Heisenberg-Ising
model in the thermodynamic limit and the lower panel the Heisenberg
model with $N=24$ spins. Lines show values of constant entropy,
starting with $S/N=0.02$ at the bottom and increasing in steps
of $0.02$.
\label{fig:entropy1} }
\end{figure}


Fig.~\ref{fig:entropy1} shows our results for the entropy per spin
$S/N$ at $J_1=K_1=1$, $J=K_2=0.7$ as a function of magnetic field
$H$ and temperature $T$. Here we focus on magnetic fields
corresponding to magnetizations $M \le 1/4$. At a quantitative level
there are certain differences between the Heisenberg-Ising model
(top panel of Fig.~\ref{fig:entropy1}) and the Heisenberg model
(bottom panel). Firstly, from the magnetization curve
Fig.~\ref{fig:magSet1}(a) we already know that the transition field
$H_c$ between the $M=0$ and $1/4$ states is shifted. Secondly, we
also know that the $M=0$ state of the Heisenberg model is
non-degenerate whereas the Heisenberg-Ising model has a macroscopic
degeneracy giving rise to a residual entropy $S/N = \ln2/4 =
0.173\ldots$, and accordingly, the values of $S$ differ in the
low-field low-temperature regime of Fig.~\ref{fig:entropy1}.
Nevertheless, the Heisenberg-Ising model and the Heisenberg model
share important qualitative features. In particular, in both cases
another local excitation comes down as $H \to H_c$. Noting that in
the \emph{Heisenberg model} these local excitations are not allowed to
occupy two consecutive blocks for $M \le 1/2$, one can count the
degeneracy at $H=H_c$ using a $2 \times 2$
transfer matrix\cite{ZhHo04,Huang,bax} and finds $S/N =
\ln((1+\sqrt{5})/2)/4 = 0.120\ldots$ in the limit $T \to 0$. A similar
$3 \times 3$ transfer-matrix procedure for the \emph{Heisenberg-Ising model}
yields $S/N = \ln(1+\sqrt{2})/4 = 0.220\ldots = \ln2/4+ 0.047\ldots$ for the
zero-temperature limit at $H=H_c$.
Recall also that the ground state at
$M=1/4$ is two-fold degenerate. For the $N=24$ system shown in the
lower panel of Fig.~\ref{fig:entropy1}, this gives rise to a
finite-size value $S/N = \ln2/24 = 0.028\ldots$ for $H>H_c$ and in
the limit $T \to 0$.

The additional residual entropy at $H=H_c$ gives rise to substantial
cooling during adiabatic (de)magnetization, as also demonstrated by the
constant entropy curves in Fig.~\ref{fig:entropy1}. In fact, if one starts
with a sufficiently low temperature at $H=0$, one reaches $T=0$ as $H$
approaches $H_c$, both in the Heisenberg model and the Heisenberg-Ising
model. Similar behavior is suggested in the two-dimensional
Shastry-Sutherland model by the flat triplet branch above the dimerized
ground state.\cite{WHO99} Indeed, there is indirect evidence for
substantial cooling in SrCu$_2$(BO$_3$)$_2$ during pulse-field
magnetization experiments.\cite{LSBHTKWU08}

\begin{figure}[tb!]
\begin{center}
\includegraphics[width=\columnwidth]{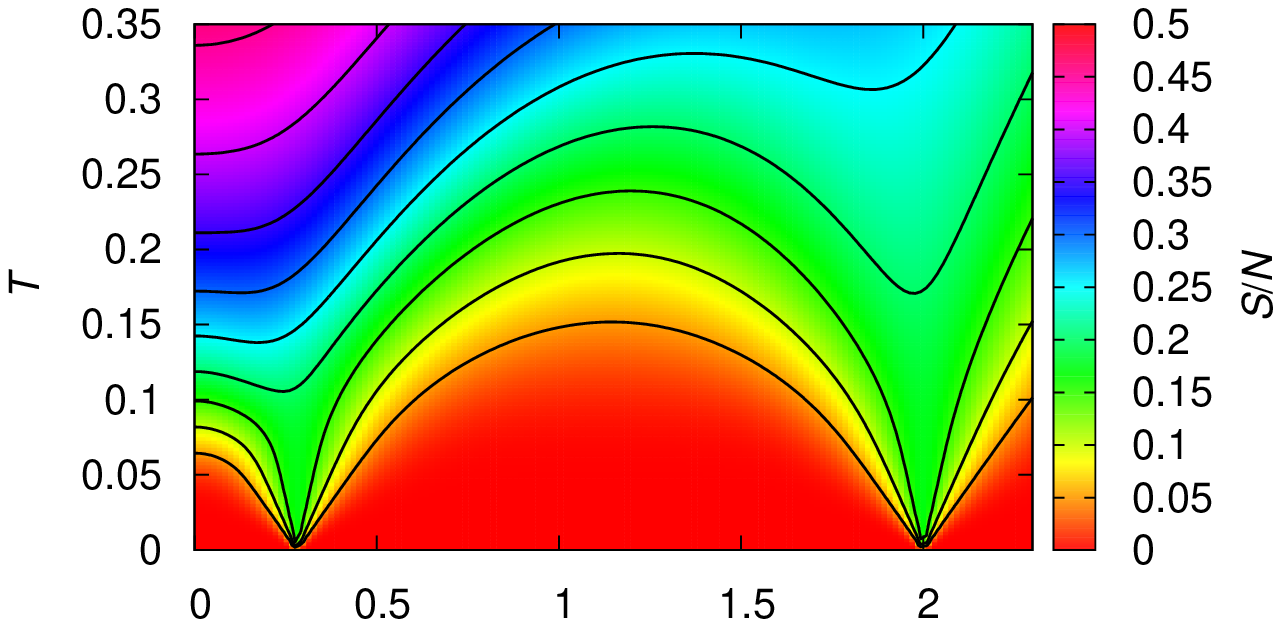}
\includegraphics[width=\columnwidth]{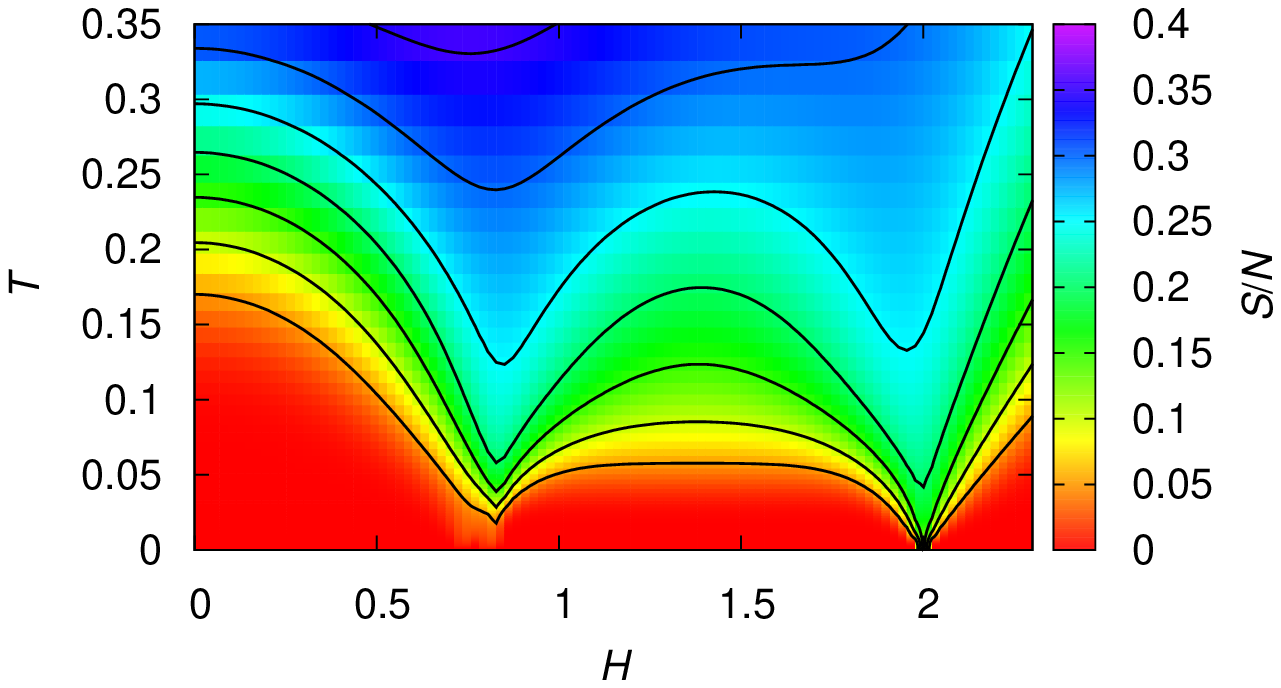}
\end{center}
\caption{(Color online) Entropy per spin $S/N$
as a function of $H$ and $T$ for $J_1=J=K_2=1$, $K_1=-1$.
The top panel shows results for the Heisenberg-Ising
model in the thermodynamic limit and the lower panel the Heisenberg
model with $N=24$ spins. Lines show values of constant entropy,
starting with $S/N=0.05$ at the bottom and increasing in steps
of $0.05$.
\label{fig:entropy2} }
\end{figure}

The other case of special interest is $J_1=J=K_2=1$, $K_1=-1$. Not only
does Fig.~\ref{fig:magSet2}(b) demonstrate quantitative agreement between
the Heisenberg model and the Heisenberg-Ising model at the saturation
field, but this case can also be considered as a realization of the
hard-monomer universality class of localized
magnons.\cite{DR04,Richter05,DeRi06,DRHS07}

Fig.~\ref{fig:entropy2} shows the entropy per site $S/N$ for the
Heisenberg-Ising model (top panel) and the Heisenberg model (lower panel)
at $J_1=J=K_2=1$, $K_1=-1$. The global behavior of both models is similar.
The main qualitative difference can be observed at small magnetic fields
when the zero-field gap is closing: the Heisenberg-Ising model has a
macroscopic ground-state degeneracy $S/N=\ln2/4$ at the position of the
first step in the magnetization curve Fig.~\ref{fig:magSet2}(b) whereas in
the Heisenberg model the degeneracy is lifted in correspondence with the
smoother transition. Nevertheless, also the Heisenberg model exhibits an
enhanced low-temperature entropy as the $T=0$ magnetization increases from
$M=0$ to $1/2$. This enhanced entropy is reflected by the minimum in the
constant entropy curves in the lower panel of Fig.~\ref{fig:entropy2}.

Close to the saturation field $H=2$ we find even quantitative agreement
between the Heisenberg-Ising and Heisenberg model. In particular, both
models now give rise to a zero-temperature entropy $S/N=\ln2/4$ at $H=2$
such that one could cool to arbitrarily low temperatures by adiabatic
magnetization or demagnetization when approaching the saturation field
from below or above, respectively. Note that in this regime finite-size
effects are very small, as is evident, e.g., from the effective
hard-monomer description.\cite{DR04,Richter05,DeRi06,DRHS07} Hence, we can
use our $N=24$ exact diagonalization data for reference.


\begin{figure}[tb!]
\begin{center}
\includegraphics[width=\columnwidth]{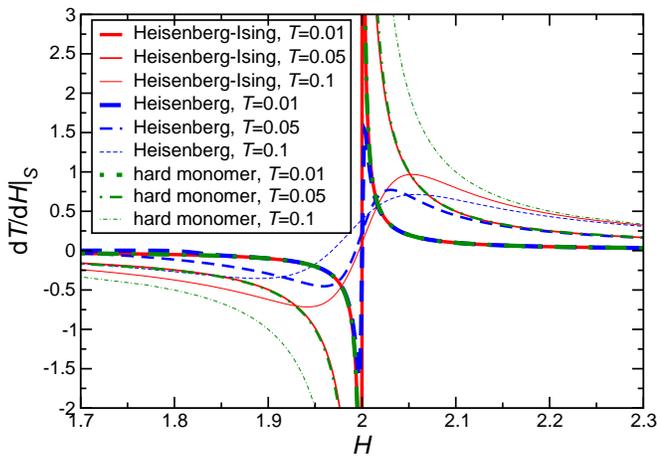}
\end{center}
\caption{(Color online) Adiabatic cooling rate as a function of
magnetic field $H$ for three different temperatures at $J_1=J=K_2=1$,
$K_1=-1$. Results for the Heisenberg-Ising model and hard monomers are for
the thermodynamic limit while those for the Heisenberg model are for
$N=24$ spins. \label{fig:dTdHcomp} }
\end{figure}

For a more detailed comparison we use the adiabatic cooling rate
which can be written as
\begin{equation}
\left.\frac{{\rm d}\, T}{{\rm d}H}\right|_S =
-\frac{{\partial\,S}/{\partial H}|_T}{{\partial\,S}/{\partial T}|_H}\,.
\label{eq:defDTDH}
\end{equation}
This is nothing else but the slope of the constant entropy curves in
Figs.~\ref{fig:entropy1} and \ref{fig:entropy2}. The adiabatic
cooling rate has the advantage over the entropy that it is directly
accessible in experiments.\cite{LTWJTHRPAD10,ToGe11,WTJTHRHPADL11}

For hard monomers, the right-hand side
of Eq.~(\ref{eq:defDTDH}) can be evaluated easily using the free energy
per unit cell $f = -T\,\ln\left(1+\exp(-(H-H_c)/T)\right)$ and
$S=-{\partial\,f}/{\partial T}$:
\begin{equation}
\left.\frac{{\rm d}\, T}{{\rm d}H}\right|_S^{\rm monomer} =
\frac{T}{H-H_c} \, .
\label{eq:dTdHmonomer}
\end{equation}
Fig.~\ref{fig:dTdHcomp} compares the cooling rate for the Heisenberg
model, the Heisenberg-Ising model, and the effective hard monomer low-energy
description for the same parameters as in Fig.~\ref{fig:entropy2}
for three different temperatures. All three descriptions exhibit a strongly
enhanced cooling rate for $H \approx H_c=2$ which reflects the
$T=0$ entropy at $H_c=2$ (see Fig.~\ref{fig:entropy2}). Not surprisingly,
the quantitative agreement between the different descriptions is best
at the lowest temperature $T=0.01$. The main difference between the
Heisenberg-Ising model and hard monomers is that the singularity
in Eq.~(\ref{eq:dTdHmonomer}) at $H=H_c$ and any $T>0$ is removed in the
Heisenberg-Ising model by the
presence of further degrees of freedom at higher energies. This improves
the agreement with the full Heisenberg model at higher temperatures.

\section{Conclusion}

We have presented a detailed analysis of the ground states and magnetic
properties of a one-dimensional lattice spin model with a structure which
is very similar to the famous orthogonal dimer
chain,\cite{odc,odc2,odc3,HSR04} although the exchange constants are more
general. In the Heisenberg-Ising variant, the main ingredient is the block
structure of the Hamiltonian. Each block consists of a triangle of $s=1/2$
spins interacting with each other via an $XXZ$-interaction and of one
single spin, which is connected to two ``dimer'' spins from the triangle
by Ising bonds. This particular structure renders the system exactly
solvable by the classical transfer-matrix method.

First, we investigated the large variety of $T=0$ ground states in an
external magnetic field. Some of the ground states break translational
symmetry spontaneously, thus giving rise to doubling of the unit cell.
Moreover, we found two macroscopically degenerate ground states. One of
them, the zero-magnetization state $|\text{AF1}\rangle$, is the closest
analog of the famous dimerized ground state of the underlying purely
quantum orthogonal dimer chain.\cite{odc,odc2,odc3} The main difference is
that the Heisenberg-Ising system cannot form a quantum dimer on the
horizontal bond, but instead the two classical antiparallel spin
configurations are equally likely to be found on each given horizontal
bond, thus yielding a two-fold degeneracy per block. The other ground
state with a two-fold degeneracy per block, $|\text{F4}\rangle$, carries
magnetization $M=1/2$ and has a different origin. It appears only at very
symmetric values of parameters and corresponds to a degeneracy between the
symmetric and antisymmetric $S_{\text{tot}}^z=1/2$ ground states for the
two coupled quantum spins on a vertical dimer.

The second focus of our investigation was a comparison of the
Heisenberg-Ising model with numerical results for the full Heisenberg
model. Certain features of the full Heisenberg model are not present in
the Heisenberg-Ising variant. For example, for certain antiferromagnetic
values of the exchange constants, the Heisenberg model also exhibits
ground states with a periodicity larger than two,\cite{odc3,SchRi02b} and
relatedly an infinite sequence of plateaux in its magnetization
curve.\cite{odc3} Only the main plateaux at $M=0$, $1/2$, and $1/4$ are
also present in the Heisenberg-Ising variant.

When considered as an approximation to the Heisenberg model, the
Heisenberg-Ising model performs in general better if some of the Ising
exchanges are ferromagnetic. In the three examples with a ferromagnetic
$K_1 < 0$ which we have discussed, the presence and nature of plateaux at
$M=0$ and $1/2$ is qualitatively reproduced to the extent possible.
Between these plateaux, the Heisenberg-Ising model always yields jumps in
the $T=0$ magnetization curve while the Heisenberg model may exhibit
smooth transitions. Also the critical fields can differ noticeably.
Remarkably, in the case $J_1=J=K_2=1$, $K_1=-1$ we found that the
Heisenberg-Ising model reproduces results for the Heisenberg model even
at a  quantitative level close to the saturation field
(see Figs.\ \ref{fig:magSet2}(b) and  \ref{fig:dTdHcomp}). This exact
correspondence may be attributed to the
product-state structure of the low-energy states of the Heisenberg model
in this parameter regime,\cite{DR04,Richter05,DeRi06,DRHS07} and we
speculate that it is generic for a sufficiently strong coupling of the
vertical dimer exchange $J_1$.

We believe that further continuation of recent developments in the field
of magnetochemistry\cite{chem,ring,Dy_10,RCRS04,JACS10,SSR11} will make it
possible to obtain magnetic materials which match the model considered in
the present paper. At the moment, the most direct application of our
system probably is as a one-dimensional version of the two-dimensional
Shastry-Sutherland model for SrCu$_2$(BO$_3$)$_2$.\cite{MiUe03} In
particular, we have observed an enhanced magnetocaloric effect upon
closing the zero-field spin gap, both in the one-dimensional Heisenberg as
well as its simplified Heisenberg-Ising variant, in accordance with
indirect evidence for cooling by adiabatic magnetization of
SrCu$_2$(BO$_3$)$_2$.\cite{LSBHTKWU08}

\acknowledgments

V.O.\ expresses his gratitude to the Institut f\"{u}r Theoretische Physik
of G\"{o}ttingen University for the warm hospitality during the course of
this project. This work was supported by the DFG (projects No.\
HO~2325/4-2, HO~2325/5-1, HO~2325/7-1, and HO~2325/8-1). V.O.\ also
acknowledges support from a joint grant of CRDF-NFSAT and the State
Committee of Science of Republic of Armenia ECSP-09-94-SASP,SCS-BFBR
11RB-001, Volks\-wagen Foundation (grant No.\ I/84 496), and
ANSEF-2497-PS.


\begin{thebibliography}{99}
\bibitem{footnote2}%
  {The numbers quoted are actually for an $N=96$ periodic
  system. However, comparison with smaller system sizes demonstrates that
  all given digits
  can be considered representative of the thermodynamic limit $N=\infty$.}%
\end{thebibliography}
\end{document}